\documentclass{article} 
\usepackage{iclr2025_conference,times}

\usepackage{amsmath,amsfonts,bm}

\def\eqref#1{equation~\ref{#1}}

\def\1{\bm{1}}

\DeclareMathAlphabet{\mathsfit}{\encodingdefault}{\sfdefault}{m}{sl}
\SetMathAlphabet{\mathsfit}{bold}{\encodingdefault}{\sfdefault}{bx}{n}

\DeclareMathOperator*{\argmax}{arg\,max}

\usepackage{hyperref}
\usepackage{url}
\usepackage{xspace}
\usepackage{bm}
\usepackage{bbm}
\usepackage{booktabs}
\usepackage{multirow}
\usepackage{wrapfig}
\usepackage{graphicx}
\usepackage{caption}
\usepackage[table]{xcolor}
\usepackage{amssymb}
\usepackage{algorithm,algpseudocode}
\usepackage{threeparttable}

\title{Tokenizing Loops of Antibodies}

\author{
\textbf{Ada Fang}$^{1,2\dagger}$ \quad
\textbf{Robert G. Alberstein}$^{2}$ \quad
\textbf{Simon Kelow}$^{2}$ \quad
\textbf{Frédéric A. Dreyer}$^{2}$ \\
$^1$Harvard University $^2$Prescient Design, Genentech \\
$^\dagger$Work done during an internship at Prescient Design, Genentech.
}

\newcommand{\xhdr}[1]{\vspace{1.7mm}\noindent{{\bf #1.}}}

\newcommand{\name}{\textsc{Igloo}\xspace}
\newcommand{\plmname}{\textsc{IglooLM}\xspace}
\newcommand{\almname}{\textsc{IglooALM}\xspace}

\iclrfinalcopy 
\begin{document}

\maketitle

\begin{abstract}
The complementarity-determining regions (CDRs) of antibodies are loop structures that are key to their interactions with antigens, and of high importance to the design of novel biologics. Since the 1980s, categorizing the diversity of CDR structures into canonical clusters has enabled the identification of key structural motifs of antibodies.  However, existing approaches have limited coverage and cannot be readily incorporated into protein foundation models. Here we introduce \textbf{I}mmuno\textbf{G}lobulin \textbf{LOO}p Tokenizer, \textbf{\name}, a multimodal antibody loop tokenizer that encodes backbone dihedral angles and sequence. \name is trained using a contrastive learning objective to map loops with similar backbone dihedral angles closer together in latent space. Compared to state-of-the-art protein encoding approaches, \name can efficiently retrieve the closest matching loop structures from a structural antibody database, outperforming the existing methods on identifying similar H3 loops by 5.9\%. \name assigns tokens to all loops, addressing the limited coverage issue of canonical clusters, while retaining the ability to recover canonical loop conformations. To demonstrate the versatility of \name tokens, we show that they can be incorporated into protein language models with \plmname and \almname. On predicting binding affinity of heavy chain variants, \plmname outperforms the base protein language model on 8 out of 10 antibody-antigen targets. Additionally, it is on par with existing state-of-the-art sequence-based and multimodal protein language models, performing comparably to models with $7\times$ more parameters. \almname samples antibody loops which are diverse in sequence and more consistent in structure than state-of-the-art antibody inverse folding models. \name demonstrates the benefit of introducing multimodal tokens for antibody loops for encoding the diverse landscape of antibody loops, improving protein foundation models, and for antibody CDR design.
\end{abstract}

\section{Introduction}
Antibodies are a class of proteins that are essential in the body's immune response and a widely used therapeutic modality~\citep{crescioli2025antibodies}. They are comprised of two identical light and two identical heavy chains. The light and heavy chains are divided into a constant and variable domain, where the variable domain is comprised of complementarity-determining regions (CDRs),\footnote{Here we consider four CDR regions. The fourth CDR is the loop joining the D and E strands adjacent to CDR1 and CDR2, which is often considered part of the framework~\citep{kelow2020hiding}.} which are structurally distinct flexible loops between antiparallel beta strands in the immunoglobulin fold. The CDRs play an essential role in the antibody's ability to recognize and bind antigens in a highly specific manner \citep{xu2000diversity}. Protein and antibody language models trained on amino acid sequence tokens have been powerful for learning evolutionary patterns that are useful for function prediction \citep{kulmanov2024protein}, sequence design \citep{zhao2025benchmark}, and variant effect prediction \citep{hie2024efficient, notin2023proteingym}. 

Recently, the development of multimodal protein language models \citep{su2023saprot,heinzinger2024prostt5,hayes2025simulating} has expanded to incorporate structure tokens in addition to sequence tokens. However, such approaches tokenize structures at the amino acid level, focus on reconstruction, and do not consider the higher-level organization and modularity of protein domains \citep{sigrist2010prosite, mistry2021pfam}. A multimodal tokenizer for antibodies should therefore consider the inherent organisation in antibody structures and sequences for effective representation learning.

Tokenization or clustering of immunoglobulin loop regions based on their dihedral backbone angles into `canonical clusters' has been adopted since \cite{chothia1987canonical}. Such a grouping of loops has been useful for understanding the structural diversity of antibodies \citep{teplyakov2016structural}, designing antibody loops with consistent structure and diverse sequences \citep{adolf2018rosettaantibodydesign}, and for studying conformational changes of antibody loops in MD simulations \citep{fernandez2020antibody, fernandez2019transitions}. These approaches are limited by their (1) limited coverage of antibody structures, with 20.3\% of all loops having no known matching canonical cluster, including 76.3\% of H3 loops (Table~\ref{tab:proportion_noise}), (2) all existing clusters only consider backbone coordinates or dihedral angles, without incorporating sequence information, and (3) existing clusters cannot be readily applied to protein language models. Thus, the tokenization of immunoglobulin loops for multimodal representation remains an open challenge.

\begin{figure}[h]
  \centering
  \vspace{-5pt}  
  \includegraphics[width=0.8\textwidth]{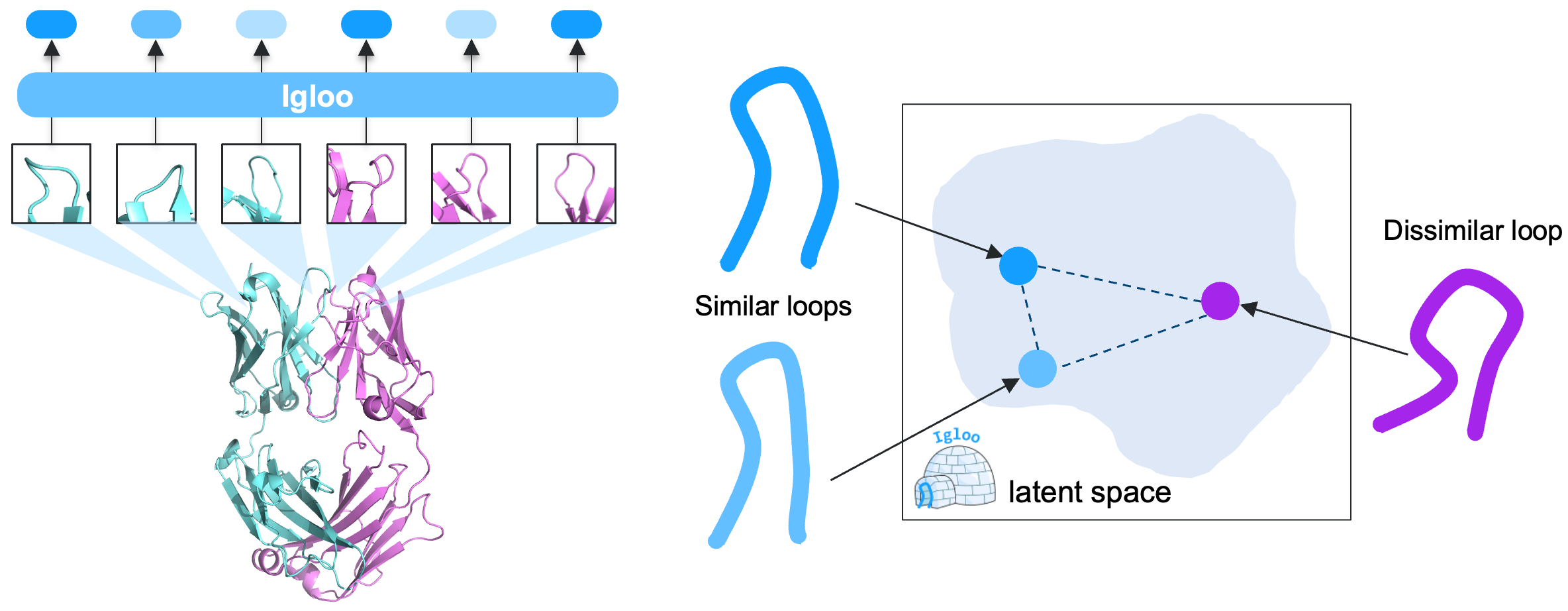}
  \vspace{-5pt}  
  \caption{\textbf{Left} \name is a multimodal tokenizer for antibody loops. \textbf{Right} Organization of the \name latent space is achieved through a contrastive learning objective on dihedral angle distance between backbones.}
  \label{fig:igloo}
\end{figure}

\xhdr{Present work} We introduce \textbf{I}mmuno\textbf{G}lobulin \textbf{LOO}p Tokenizer, \textbf{\name}, a \emph{multimodal antibody loop tokenizer} for encoding backbone dihedral angles and sequence (Fig. \ref{fig:igloo}). Unlike existing structure tokenizers, which focus on the amino acid scale, \name tokenizes at the substructure loop level. \name is trained on 807,815 loop regions from heavy and light chains of experimentally-derived and computationally predicted structures. We develop a contrastive learning objective based on the dihedral angle distance defined by \cite{north2011new} to train \name. While being a self-supervised model, \name successfully reproduces known canonical conformations assigned for 90.6\% of loops in SAbDab \cite{dunbar2014sabdab}. To demonstrate the versatility and utility of \name tokens, we present three key applications: 
\begin{itemize}
    \item \textbf{Retrieval of similar loop structures from large structural databases.} By learning to compare loop structures, \name retrieves more similar loop structures from SAbDab compared to state-of-the-art protein encoding approaches. For the H3 loop, which exhibits the most sequence and structure diversity, \name outperforms the previous best model on retrieving H3 loops with similar dihedral angle backbones by 5.9\%.
    \item \textbf{Improved antibody affinity prediction with protein language models.} We incorporate \name loop tokens into an antibody language model, and train \plmname. By using the representations learned from \plmname to predict binding affinity of heavy chain variants, we show it outperforms the base model on 8 out of 10 antibody-antigen targets and performs on par to models with $7\times$ more parameters. 
    \item \textbf{Sampling diverse loops with consistent structure.} \almname is a protein language model with the \name loop tokens and \name multimodal residue tokens. When loop sequences are masked out, the loops sampled from \almname are diverse in sequence and more consistent in structure than state-of-the-art antibody inverse folding models. Redesigned CDR H3 loops of a SARS-CoV-2 antibody with \almname achieves an average sequence identity of 0.27 while achieving less than 1\AA\ RMSD to the original loop.
\end{itemize}

By introducing multimodal tokens for antibody loops, \name captures the structural and functional diversity of loop conformations, improving the expressiveness of protein foundation models and advancing rational antibody design.

\section{Related Work}
\xhdr{Tokenization of protein structure} The construction of classifications of protein structures at the domain level has been applied for understanding the relationship with domain function \citep{lo2000scop, ouzounis2003classification, sigrist2010prosite, mistry2021pfam}. Learning structurally informed residue-level representations can be achieved with geometric features \citep{jing2020learning}, multiview contrastive learning between sequence and structure views of the same protein \citep{zhang2022protein}, hierarchical graph neural network on the protein structure \citep{wang2022learning}, and with intermolecular interactions \citep{fang2025}. Tokenization at the amino acid level has shown significant advances in the speed of protein structure search with Foldseek \citep{van2024foldseek}. \cite{yuan2025protein} compare different approaches for tokenizing amino acid structures including VQVAE \citep{hayes2025simulating} and inverse-folding-based methods \citep{proteinmpnn}. 

\xhdr{Multimodal protein language models}
Multimodal protein language models have been trained to learn meaningful representations and to generate over sequence, structure, and function. Models such as SaProt \citep{su2023saprot} and ProstT5 \citep{heinzinger2024prostt5} learn representations from protein sequence and Foldseek 3Di amino acid tokens, which capture structural information. ProSST \citep{li2024prosst} represents proteins with sequence and residue-level structure tokens that capture local environments. ProSSN \citep{tan2025semantical} uses both sequence and the topological structure of proteins to learn multimodal representations. ESM3 \citep{hayes2025simulating} is a generative model that models the sequence, structure, and function of amino acids simultaneously. 


\xhdr{Clustering Immunoglobulin Loops}
The CDRs of antibodies demonstrate the most variability and are essential to the binding of antibodies to antigens. Thus, there has been significant effort in categorizing all known structures of CDRs \citep{chothia1987canonical, shirai1996structural, north2011new, adolf2015pyigclassify, nowak2016length, wong2019comparative, kelow2022penultimate, liu2024antibody}. CDRs fold into a loop structure, and a pair of loops can be compared through their backbone dihedral angles \citep{north2011new}. While most approaches only cluster loops of the same length, \cite{nowak2016length} explore clustering of loops of different lengths by aligning loops with their stem region (subsequence of amino acids before and after the loop region) and comparing the resultant RMSD between the loops. 
SCALOP \citep{wong2019scalop} predicts canonical loops from protein sequences for large-scale annotation of antibody libraries. 
\cite{zhang2025fast} train their model to learn the RMSD between pairs of loops and show how it can be used for designing CDRs. Current methods are limited as many CDRs, especially H3 loops, do not have known canonical conformations (Table~\ref{tab:proportion_noise}). We extend existing approaches through the self-supervised definition of antibody loop clusters.

\section{Method}

\name is a multimodal tokenizer that incorporates both sequence and backbone structure of the loop structures. Here we focus on modeling loops within antibodies and TCRs, which are the four CDRs of the heavy and light chains. \name is a \textit{tokenizing function} that maps for a loop sequence and backbone structure, to a token $\mathbf{t}$.

\begin{figure}[H]
  \centering
  \vspace{-10pt}  
  \includegraphics[width=0.4\textwidth]{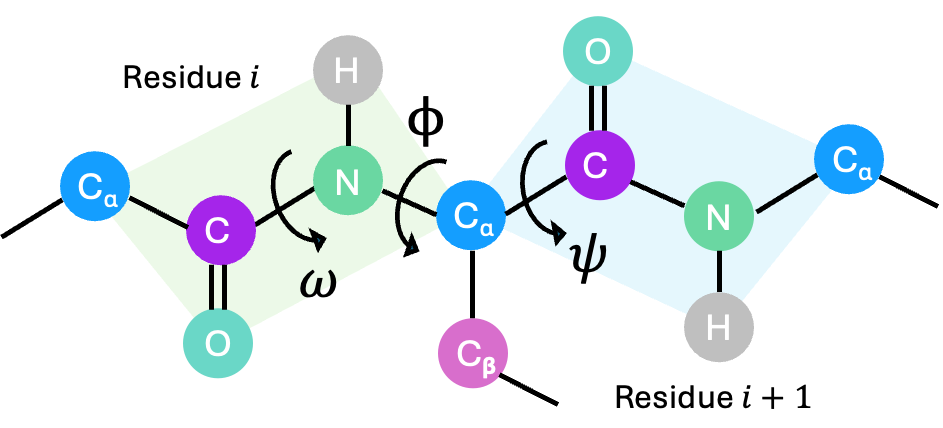}
  \vspace{-10pt}  
  \caption{Backbone dihedral angles for residue $i$.}
  \vspace{-10pt}  
  \label{fig:dihedral_angles}
\end{figure}

\xhdr{Problem definition} An antibody loop with $n$ residues is defined by: (1) a sequence of amino acids $\mathbf{a}=(a_1,\dots, a_n)$ where $\forall\, i,\, a_i \in \mathcal{V}=\{\text{Ala}, \text{Arg}, \dots, \text{Tyr}, \text{Val}\}$, which are canonical amino acid residues, and (2) their backbone dihedral angles $\bm{\phi}, \bm{\psi}, \bm{\omega} \in (-\pi,\pi]^{n}$ (Fig.~\ref{fig:dihedral_angles}). Our goal is to train a tokenizer $f(\cdot)$ for antibody loops such that $f(\mathbf{a}, \bm{\phi}, \bm{\psi}, \bm{\omega})=\mathbf{t}$, where $\mathbf{t} \in \mathbb{R}^d$ and $d$ are the dimensions of the token embeddings. For the token $\mathbf{t}$, it can be used to determine loops that share a similar structure, incorporation into a protein language model, and for guided generation.

\subsection{Multimodal Tokenization of Loops}
The input to \name is a loop of length $n$ with dihedral angles $(\bm{\phi}, \bm{\omega}, \bm{\psi}) \in (-\pi, \pi]^{n\times3}$ and amino acid identities $\mathbf{a}$. The dihedral angles are first converted into coordinates on the unit circle $(\cos\bm{\phi}, \sin\bm{\phi}, \cos\bm{\psi}, \sin\bm{\psi}, \cos\bm{\omega}, \sin\bm{\omega}) \in [-1,1]^{n\times6}$ and then projected with a linear layer $\mathbf{D} = ( \cos\bm{\phi}, \sin\bm{\phi}, \cos\bm{\psi}, \sin\bm{\psi}, \cos\bm{\omega}, \sin\bm{\omega}) \mathbf{W_{\text{dihedral}}}^T + \mathbf{b_\text{dihedral}}$, where $\mathbf{D} \in \mathbb{R}^{n\times d}, \mathbf{W_{\text{dihedral}}} \in \mathbb{R}^{d\times 6}, \text{ and } \mathbf{b_\text{dihedral}} \in \mathbb{R}^d$. For the loop sequence, it is encoded with 20 learnable embeddings for each of the canonical amino acid types into $\mathbf{A} \in \mathbb{R}^{n\times d}$. Next, we sum the sequence and dihedral angle embeddings to produce a multimodal embedding, $\mathbf{X} =\mathbf{D} + \mathbf{A} \in \mathbb{R}^{n\times d}$ (Fig.~\ref{fig:igloo_model}a). To learn a representation across the loop residues $\mathbf{X}=(\mathbf{x}_1, \dots, \mathbf{x}_n)$, we use a transformer architecture based on BERT \citep{bert} using the ESM-2 implementation \citep{lin2023esm2}. A learnable classification token, $\mathbf{t}$, is added to the start of each sequence to learn a meaningful overall representation of the loop.

\subsection{\name Self-Supervised Training Objectives}\label{sec:igloo-objective}

We train \name with four objective functions (Fig.~\ref{fig:igloo_model}b): (1) masked reconstruction of dihedral angles, (2) masked reconstruction of amino acid identities, (3) contrastive learning of protein backbones, and (4) codebook learning. Following the multimodal masking approach of ESM-3 \citep{hayes2025simulating}, we randomly mask 30\% of the input tokens and vary the masking patterns across tracks as described below, ensuring that \name is exposed to every possible combination of tracks during reconstruction.

\xhdr{Masked reconstruction of dihedral angles} For masking of sequence tokens, the following approaches are used (1) Masking uniformly at random the dihedral angles and protein sequence for 20\% of inputs, (2) masking uniformly at random only the dihedral angles for 20\% of inputs, (3) completely masking the dihedral angles, keeping only the protein sequence for 10\% of inputs. For dihedral angle $\theta_{i}$ of residue $i$ in a loop, the predicted unit circle coordinates $(x_{i},y_{i})$ are given by passing the hidden representation of the residue to a two-layer MLP. The reconstruction loss ${\ell_{\text{dihedral recon.}}}_i$ is given by the mean squared error between $(\cos\theta_{i}, \sin\theta_{i})$ and  $(\cos\hat\theta_{i}, \sin\hat\theta_{i})$, where $\hat\theta_{i} = \arctan (y_{i}/x_{i})$. We also add a penalty term, ${\ell_{\text{dihedral reg.}}}$, to regularize the model for the reconstruction of coordinates on the unit circle \citep{pavllo2018quaternet}.

\xhdr{Masked reconstruction of amino acid identities} We employ several variations of masking tokens across the inputs for 20\% of inputs. (1) Masking uniformly at random the protein sequence and the dihedral angles, (2) masking uniformly at random only the protein sequence for 20\% of inputs, (3) completely masking the protein sequence, keeping only the dihedral angles for 10\% of inputs. The logits of the masked amino acids are given by passing the hidden representation of the residue to a two-layer MLP. The amino acid masking loss for amino acid $i$ in a loop, ${\ell_{\text{AA}, i}}$, is given by the cross-entropy loss of the predicted amino acid identity and the true amino acid identity. 

\xhdr{Contrastive learning of protein backbones} For \name to learn a token, $\textbf{t}$, such that similar tokens share similar loop conformations, we define a contrastive loss function. \cite{north2011new} defines the similarity between two loops $u$ and $v$ of length $n_u$ and $n_v$, respectively, and $n_u \leq n_v$. The loops have dihedral angles $( \bm{\phi}^u,  \bm{\psi}^u, \bm{\omega}^u) \in (-\pi,\pi]^{n_u\times 3}$ and $(\bm{\phi}^v, \bm{\psi}^v, \bm{\omega}^v) \in (-\pi,\pi]^{n_v\times 3}$. The dihedral distance $\mathcal{D}$ is defined as
\begin{equation}\label{eqn:dihedral}
\mathcal{D}=\frac{1}{3M}\sum_{\theta\in\{\phi,\psi,\omega\}}\sum_{i=1}^{n_u} 2(1-\cos\bigr(\theta_{i}^u-\mathcal{P}(\theta_{i}^v))\bigl), 
\end{equation}
where $\mathcal{P}$ aligns residues of loop $v$ to residues of loop $u$. When $n_u = n_v$, the alignment is a one-to-one mapping between the residues. Otherwise, we define an alignment between the two loops using a dynamic time warping path (see Appendix~\ref{sec:loop-length-independent}).

Proteins are chiral molecules, and the orientation of the backbone frame has a strong influence on the atomic structure through side-chain positioning. Prominent examples exist where the                                                       backbone RMSD of a pair of loops is low, yet dihedral angles can be up to $180^\circ$ apart with opposite-pointing side chains \citep{north2011new}. Therefore, we use dihedral angle distance $\mathcal{D}$ over RMSD to capture the nuances of the loop structure.

A pair of loops $u,v$ is a positive pair ($Y_{uv} = 1$) if the loops are of the same length and $\mathcal{D} < 0.1$. A pair of loops is a negative pair ($Y_{uv} = 0$) if they are of different lengths or $\mathcal{D} > 0.47$ for loops of the same length, where $\mathcal{D} = 0.47$ corresponds to an average difference in dihedral angles of $40^\circ$, which is the threshold used in the clustering by \cite{kelow2022penultimate}. Otherwise, the pair of loops is ignored. The \emph{dihedral loss} is the mean binary cross-entropy over the pairs of loops in the batch with positive and negative labels.
\begin{equation}
   \ell_{\text{contrastive},uv}
      \;=\;
      \mathrm{BCE}\!\Bigl(\sigma\Bigl(\frac{\mathbf{h}_u^{\!\top}\mathbf{h}_v}{\tau}\Bigr),\,Y_{uv}\Bigr),
\end{equation}
where $\mathbf{h}_u= \frac{\mathbf{t}_u}{\lVert\mathbf{t}_u\rVert_2}$, $\mathbf{t}_u$ is the classification token embedding for loop $u$ in the batch, and $\tau$ is the temperature. We apply contrastive learning instead of predicting $\mathcal{D}$ as the pretraining task for similar loops to be close in the latent space. A margin between positive and negative pairs is applied so \name does not overfit its representations to the threshold used for the definition of canonical clusters.


\xhdr{Codebook learning} In addition to learning continuous tokens, the assignment of loops to $K$ discrete tokens offers a convenient and fast approach for loop comparison required for high-throughput queries. For the codebook $C \in \mathbb{R}^{K\times d}$ to learn quantized tokens $\mathbf{\hat{t}}$, we include a codebook learning loss \citep{van2017neural} on the classification token of loop $u$ with $\ell_{\text{codebook, }u} = ||\,\text{sg}[\mathbf{t}_u] - \mathbf{\hat{t}}_u\,||^2_2 + \alpha  ||\,\mathbf{t}_u - \text{sg}[\mathbf{\hat{t}}_u] \,||^2_2$, where \text{sg} is the stop gradient operator and $\alpha$ is the weight on the second commitment loss term. 

\subsection{Training and Inference of \name}
For the training of \name we use the overall loss function, which is given by
\begin{equation}
    \mathcal{L} = \ell_{\text{dihedral recon.}} + \ell_{\text{AA}} + \ell_{\text{contrastive}} + \ell_{\text{codebook}} + \lambda {\ell_{\text{dihedral reg.}}}.
\end{equation}
We train \name on heavy and light chain CDR1, CDR2, CDR3, and CDR4 loops from all antibodies and nanobodies in SAbDab \citep{dunbar2014sabdab}, and TCRs in STCRDab \citep{leem2018stcrdab}. In addition, we also train with Ibex \citep{ibex} predicted structures of paired heavy and light chain antibodies from paired sequences of the Observed Antibody Space (OAS, \cite{olsen2022oas}) (Appendix~\ref{appendix:data-igloo}). For each antibody, we then use their concatenated CDR sequence and an 80\% sequence identity threshold for splitting loops of antibodies into train, test, and validation. Since clustering at the level of concatenated sequences of CDRs can still result in an individual CDR sharing the same sequence, any training loop sequences are removed from the validation and test set. In total, \name is trained on 108,167 experimentally resolved loop structures from SAbDab and STCRDab and 699,648 predicted loop structures from paired OAS sequences. At inference, \name outputs a continuous classification loop token $\mathbf{t}$, a quantized token $\hat{\mathbf{t}}$, and a multimodal representation for each residue $i$ in the loop $\mathbf{x}_i$. 

\subsection{Incorporating \name tokens into protein language models}
\begin{figure}[h]
  \centering
  \vspace{-10pt}  
  \includegraphics[width=\textwidth]{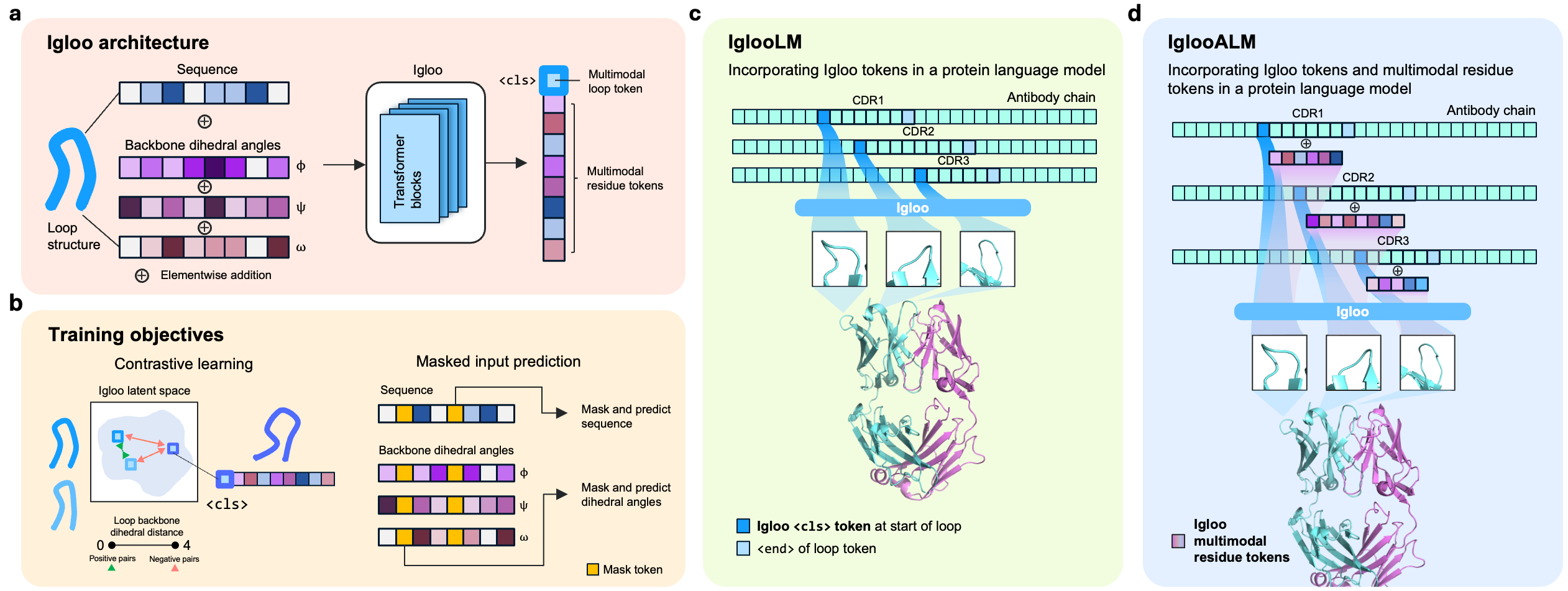}
  \vspace{-10pt}  
  \caption{
  \textbf{a} \name is a multimodal tokenizer for antibody loops.
  \textbf{b} Training objectives involve (1) contrastive learning with positive and negative pairs defined by their dihedral angle distance, and (2) masking and prediction of sequence and backbone dihedral angles.
  How \name tokens of CDR loops can be incorporated into protein language models where \textbf{c} \plmname contains only the \texttt{<cls>} (classification) loop token, $\mathbf{t}$, and \textbf{d} \almname contains the loop token and multimodal residue tokens.
  }
  \vspace{-5mm}
  \label{fig:igloo_model}
\end{figure}

\xhdr{Approach} We demonstrate how \name loop tokens, $\mathbf{t}$, can be inserted as special tokens in protein language models with two complementary approaches. (1) \plmname (Fig.~\ref{fig:igloo_model}c) is a protein language model with the \name loop token, $\mathbf{t}$, inserted at the start of each CDR loop and an \texttt{<end>} token added at the end of the loop. (2) \almname (ALM=dihedral Angle Language Model, Fig.~\ref{fig:igloo_model}d) is a protein language model with the \name loop token, $\mathbf{t}$, and \name multimodal residue tokens, $\mathbf{x}_i$, for each amino acid in the CDR loop. These models are finetuned from the 420M parameter base antibody language model, IgBert \citep{kenlay2024igbert}, a BERT-style model trained on all paired and unpaired OAS sequences. We project \name tokens with a linear layer so that they are the same dimension as the hidden dimension of tokens in the base protein language model.

Learnable classification tokens have been widely used in text, vision, and single-cell transformers \citep{bert, dosovitskiy2020image, cui2024scgpt}. Analogous to the cell-prompting and gene-prompting paradigms of scGPT \citep{cui2024scgpt}, \plmname encodes loops with tokens, $\mathbf{t}$, while \almname extends this scheme by combining loop tokens $\mathbf{t}$ and multimodal residue tokens, $\mathbf{x}$. Embeddings from \plmname contain the context of the loop conformation, while embeddings from \almname additionally contain the context of the dihedral angles of each residue in the loop. We demonstrate how \almname excels in tasks where the accurate residue-level structure is provided. Conversely, \plmname excels in tasks where the loop conformation is known, but accurate residue-level structure prediction is challenging -- for example, deep-mutational-scan datasets in which sequences differ by only a few point mutations \citep{pak2023using, buel2022can}. 


\xhdr{Training and Inference} The models are finetuned with the same masked language model objective as \cite{kenlay2024igbert}--uniformly randomly masking 15\% of amino acid residues, for which 80\% are then replaced by a masked token, 10\% are changed to a random token in the vocabulary, and 10\% are left unchanged. The same residues are also masked for the computation of the \name token. \plmname and \almname are trained on single domains from paired OAS sequences, which are split into train, validation, and test splits based on a 90\% sequence identity split (Appendix~\ref{appendix:data-plm}). All structures of the loops required for \name tokens are extracted from Ibex predicted structures.

\section{Experiments}\label{sec:experiments}

\subsection{\name for Paratope Retrieval}\label{exp:paratope}
In this evaluation, for a set of query CDRs, we task \name to retrieve from a large repository of CDRs those with the closest experimentally determined backbone structure, thereby directly assessing how well the representation captures paratope-level structure.

\xhdr{Experimental set up} Query CDRs are from the \name unseen test set of CDRs, and the repository CDRs are those from the train and validation set of SAbDab. We use the \name $\mathbf{t}$ token and retrieve 20 loops with the highest cosine similarity from loops of the same type and length. Retrieved CDRs are deemed correct if $\mathcal{D} < 0.47$ (Eq. \ref{eqn:dihedral}) or $\text{RMSD}<1$ \AA\ to the query CDR loop. 

\xhdr{Baselines} We compare \name to protein language models that are trained on UniProt: ESM C \citep{ESMCambrian2024} and ESM-2 (3B) \citep{lin2023esm2}, and models trained on OAS: AbLang2 \citep{olsen2024addressing} and IgBert \citep{kenlay2024igbert}. Additionally, we evaluate the quality of retrieval when structure is also given and compare to multimodal protein language models, SaProt \citep{su2023saprot} and ProstT5 \citep{heinzinger2024prostt5}, which also take as input the Foldseek 3Di tokens \citep{van2024foldseek} derived from the protein structure. To ensure embeddings focus on the loop, for models which embed the whole protein sequence, the loop embedding is defined as the mean embedding over the amino acids in the loop. We also compare to continuous structure tokenizers which were benchmarked by \citep{yuan2025protein}. This includes inverse folding models: MIF \citep{yang2023masked} and ProteinMPNN \citep{proteinmpnn}, and the continuous encoder embedding of VQVAE models: Foldseek 3Di and Amino Aseed \citep{yuan2025protein}. For further details see Appendix \ref{appendix:retrieval-method}.

\xhdr{Results} We evaluate the models with precision at rank 20 and observe that \name achieves state-of-the-art performance in retrieving similar paratopes, based on dihedral distance $\mathcal{D} < 0.47$ and $\text{RMSD}<1$ \AA\ from loops of the same length when given sequence and dihedral angle input (Table~\ref{tab:retrieval}). Compared to larger protein language models pretrained with masked language modeling, \name and Foldseek achieve higher precision. For the retrieval of similar H3 loop structures based on dihedral distance $\mathcal{D}$, \name outperforms the best structure tokenizer, Amino Aseed, by 5.9\% and the best protein language model or antibody language model, ESM-2 (3B) by 69.8\%. The H3 loop is particularly hard to represent for sequence-only language models due to the high sequence diversity owing to V(D)J recombination \citep{tonegawa1983somatic}. We also show how ablating key components of the model effects performance in Appendix~\ref{sec:ablation}.

\begin{table}[h]
\centering
\caption{Average precision at rank 20 for retrieval of similar CDR paratopes.  
Models are shown in rows.  The \textbf{first}, \underline{second}, and \textit{third} best performance for each column are highlighted. Additional results for precision at rank 1, 5, and 10 are available at Table~\ref{tab:retrieval-topk}.}
\tiny
\vspace{1mm}
\label{tab:retrieval}
\begin{tabular}{ll|cccccc|cccccc}
\toprule
 &
& \multicolumn{6}{c|}{\textbf{\% RMSD $<$ 1~\AA}} &
\multicolumn{6}{c}{$\%\,\mathbf{\mathcal{D}<0.47}$} \\
 \multicolumn{2}{c|}{\textbf{Model}} &
\textbf{L1} & \textbf{L2} & \textbf{L3} & \textbf{H1} & \textbf{H2} & \textbf{H3} & \textbf{L1} & \textbf{L2} & \textbf{L3} & \textbf{H1} & \textbf{H2} & \textbf{H3} \\
\bottomrule
\multicolumn{2}{l|}{\textbf{Random}} & 0.545 & 0.557 & 0.373 & 0.249 & 0.351 & 0.127 & 0.648 & 0.730 & 0.392 & 0.559 & 0.508 & 0.126 \\
\midrule
\multirow{2}{*}{\textbf{PLM}} & \textbf{ESM C}          & 0.750 & 0.700 & 0.489 & 0.418 & 0.519 & 0.19 & 0.811 & 0.916 & 0.517 & 0.692 & 0.702 & 0.208 \\
 & \textbf{ESM-2 (3B)}     & 0.740 & 0.704 & 0.500 & 0.425 & 0.522 & 0.206 & 0.802 & 0.904 & 0.534 & 0.706 & 0.688 & 0.237 \\
\midrule
\multirow{2}{*}{\textbf{AbLM}} & \textbf{AbLang2}        & 0.689 & 0.604 & 0.482 & 0.402 & 0.497 & 0.173 & 0.761 & 0.782 & 0.537 & 0.602 & 0.699 & 0.222 \\
& \textbf{IgBert}         & 0.705 & 0.622 & 0.482 & 0.377 & 0.479 & 0.182 & 0.773 & 0.813 & 0.511 & 0.709 & 0.677 & 0.216 \\
\midrule
\multirow{2}{*}{\textbf{MPLM}} & \textbf{SaProt}         & 0.737 & 0.704 & 0.499 & 0.420 & 0.491 & 0.218 & 0.790 & 0.918 & 0.578 & 0.688 & 0.646 & 0.248 \\
& \textbf{ProstT5}        & 0.782 & \textbf{0.716} & 0.539 & 0.458 & \underline{0.586} & 0.276 & 0.846 & 0.941 & \textit{0.629} & 0.711 & \textbf{0.756} & 0.359 \\
\midrule
\multirow{2}{*}{\textbf{IF}}  & \textbf{MIF}            & 0.776 & 0.699 & 0.516 & 0.432 & 0.491 & 0.231 & 0.833 & 0.933 & 0.604 & 0.702 & 0.641 & 0.298 \\
& \textbf{ProteinMPNN}    & \underline{0.804} & 0.700 & \textit{0.546} & \underline{0.459} & 0.521 & \underline{0.286} & 0.839 & \textit{0.943} & 0.632 & \textbf{0.732} & 0.710 & \textit{0.372} \\
\midrule
\multirow{2}{*}{\textbf{VQVAE}}  & \textbf{Foldseek 3Di}       & 0.785 & 0.696 & \underline{0.556} & \textbf{0.467} & \textbf{0.591} & \textit{0.281} & \underline{0.849} & 0.909 & \underline{0.640} & \underline{0.715} & \textit{0.730} & 0.362 \\
& \textbf{Amino Aseed}    & \textbf{0.812} & \underline{0.713} & 0.542 & 0.420 & 0.529 & \textbf{0.292} & \textbf{0.851} & \underline{0.952} & 0.625 & 0.688 & 0.713 & \underline{0.379} \\
\bottomrule
\rowcolor{gray!15} \textbf{Ours} & \textbf{\name} & \textit{0.793} & \textit{0.705} & \textbf{0.558} & \underline{0.459} & \textit{0.578} & 0.278 & \textbf{0.851} & \textbf{0.956} & \textbf{0.674} & \underline{0.715} & \underline{0.749} & \textbf{0.402} \\
\bottomrule
\end{tabular}
\par\vspace{2pt}
  {\scriptsize PLM: Protein Language Model, AbLM: Antibody Language Model, \\ MPLM: Multimodal Protein Language Model, IF: Inverse Folding Model}  
\vspace{-3mm}
\end{table}

\subsection{\name for Clustering Antibody Structures}
\begin{wrapfigure}{r}{0.3\textwidth}  
\vspace{-5mm}  
\centering
\captionof{table}{Average \name cluster purity ($\uparrow$) of \cite{kelow2022penultimate} defined clusters of antibody CDRs across SAbDab.}
\vspace{-2mm}
\small
\begin{tabular}{lcc}
\toprule
\textbf{CDR} & \textbf{Heavy} & \textbf{Light} \\
\midrule
CDR1 & 0.894 & 0.880 \\
CDR2 & 0.900 & 0.975 \\
CDR3 & 0.754 & 0.831 \\
CDR4 & 0.983 & 0.930 \\
\bottomrule
\end{tabular}
\vspace{-2mm}
\label{tab:codebook_purity}
\end{wrapfigure}

The canonical clusters established by \cite{north2011new} and \cite{kelow2022penultimate} have been widely used for categorizing new structures \citep{teplyakov2016structural} and to analyze
molecular dynamics simulations of antibodies \citep{fernandez2020antibody, fernandez2019transitions}. In this section, we evaluate how well the quantized token, $\hat{\mathbf{t}}$, recovers their clusters of antibody CDRs. 

\xhdr{Evaluation setup}
Let the \name learned codebook $\hat{\mathbf{t}}$ induce the partition $\mathcal{C}=\{C_{1},\ldots,C_{K}\}$, and let $\mathcal{G}=\{G_{1},\ldots,G_{L}\}$ denote the reference clusters of \cite{kelow2022penultimate}. We quantify the agreement between these two partitions using \textit{cluster purity}. For each predicted cluster $C_{k}$, we select the dominant reference class based on majority vote: $y^{\star}(k) \;=\;\argmax_{\,\ell}\;\bigl|C_{k}\cap G_{\ell}\bigr|$.
Items in $C_{k}$ whose reference label equals $y^{\star}(k)$ are considered correctly assigned. Overall accuracy is the proportion of correctly
assigned instances
\begin{equation}
  \mathrm{Purity}(\mathcal{C},\mathcal{G})
  \;=\;
  \frac{1}{N}\sum_{k=1}^{K}
    \max_{\ell}\bigl|C_{k}\cap G_{\ell}\bigr|,
  \quad N = \sum_{k=1}^{K}\lvert C_{k}\rvert .
  \label{eq:purity}
\end{equation}
We evaluate on all loops in SAbDab that can be assigned to a reference cluster with a cutoff of $\mathcal{D}=0.47$ to the centroid. A limitation of the existing canonical clustering approach is that several loops are not assigned to any cluster. We do not evaluate cluster purity on unassigned loops, which are typically referred to as belonging to ``noise'' clusters.

The reference definition of clusters assigns different clusters for different loop types and lengths. We also evaluate the cluster's \emph{loop‑type purity} and \emph{loop‑length purity} are defined as follows: 
\begin{equation}
  p^{\mathrm{type}}_k =\frac{1}{n_k}\max_{t} \sum_{x\in C_k}\mathbbm{1}\{\text{loop type}(x)=t\}, \;
  p^{\mathrm{len}}_k =\frac{1}{n_k}\max_{\ell} \sum_{x\in C_k}\mathbbm{1}\{\text{loop length}(x)=\ell\},
\end{equation}
where $t \in \{\text{H1, H2, H3, H4, L1, L2, L3, L4}\}$. We report global scores with a weighted average of the cluster-level purity scores, $P^{\mathrm{type}}=\frac{1}{N}\sum_{k} n_k\,p^{\mathrm{type}}_k, P^{\mathrm{len}}=\frac{1}{N}\sum_{k} n_k\,p^{\mathrm{len}}_k$, where $n_k$ is the number of loops in the cluster and $N=\sum_{k} n_k$.

\xhdr{Results}
Across SAbDab, 1305 \name codebooks and 180 reference clusters are used. Without exposure to loop‑type annotations, the \name{}‑induced partition is highly homogeneous, attaining a loop‑type purity of $P^{\mathrm{type}}=0.983$, and loop length purity $P^{\mathrm{len}}=0.965$. Visualization of the latent space in 2D with UMAP also shows localization of loops by loop type, length, and canonical cluster (Fig.~\ref{fig:umap}). We report cluster purity in Table \ref{tab:codebook_purity}. Our results are comparable with \cite{wong2019scalop} 
(Table~\ref{tab:sabdab_canonical_cluster_recovery}), which uses Position-Specific Scoring Matrices to predict canonical forms from sequence. These results highlight \name can recover the known canonical clusters with high purity. 

\begin{figure}[h]
\vspace{-2mm}
  \centering
  \includegraphics[width=\textwidth]{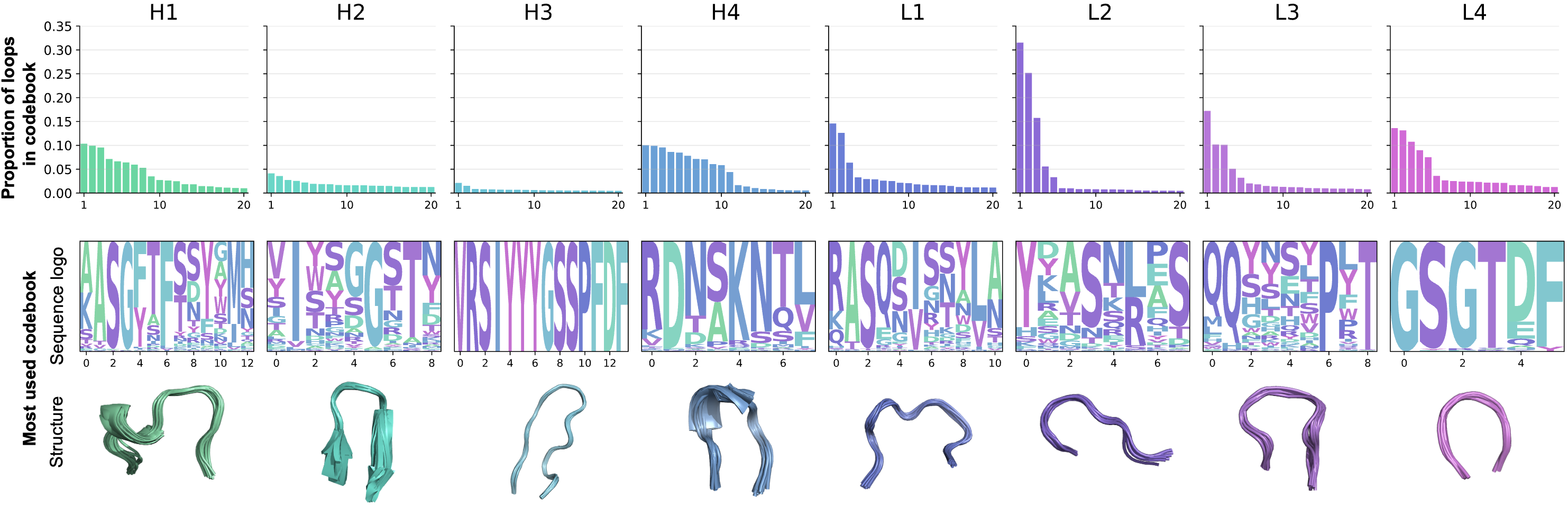}
\vspace{-4mm}
  \caption{\textbf{Top} Top 20 used \name codebooks for each CDR type in SAbDAb. \textbf{Bottom} Sequence logo and aligned structures of 20 loops for the most used \name codebook for each loop type.}
  \label{fig:loop_distribution}
\end{figure}

We further explore how different loops differ in their distribution across codebooks. The proportion of each loop in SAbDab assigned to the top 20 used codebooks is shown in Fig.~\ref{fig:loop_distribution} for each loop type. The H4 and L2 loop types have relatively low diversity with 93.0\% and 91.7\% of loops assigned to a codebook in the top 20, respectively. Conversely, H3 has the lowest coverage in the top 20 codebooks with 14.6\% of loops. The most frequent H3‑loop codebook entry appears 387 times. Every occurrence shares an identical loop sequence derived from single‑chain Fv16 antibody structures, a scaffold that is widely represented in the PDB.

\vspace{-2mm}
\subsection{Predicting Binding Affinity with \plmname}
\vspace{-2mm}

Next, \plmname is evaluated on datasets where sequences differ by a few variants. For a set of heavy chain antibody mutants, we apply the protein-level representations of the heavy chain sequences to predict binding affinity. This section aims to test if incorporating \name tokens, $\mathbf{t}$, into protein language models as a special token is beneficial to the representations learned by the model.

\xhdr{Experimental setup} We use the curated set of antibody-antigen binding affinity dataset from AbBiBench \citep{abibench}. Antibody structures for the variants are predicted with Ibex, used as input to obtain \name tokens, and embeddings are then generated with \plmname. Sequence-level embeddings are obtained by averaging residue-level embeddings. A separate model is trained for each antibody-antigen pair. Sequence-level embeddings are used as input to train a ridge regressor evaluated with 10-fold nested cross-validation (Appendix~\ref{appendix:abbibench-method}). Models are evaluated with the Spearman correlation coefficient, $\rho$, between the predicted and true binding affinity.

\xhdr{Baselines} The protein language models, ESM C, ESM-2 (3B), AbLang2, and IgBert, and multimodal protein language models SaProt and ProstT5 introduced in Section~\ref{exp:paratope} are used as baselines. We obtain sequence-level embeddings by averaging the residue-level embeddings. For the multimodal protein language models, we use Foldseek 3Di tokens from the Ibex predicted structures.

\xhdr{Results} Across the 10 antibody–antigen pairs in Table \ref{tab:abibench}, \plmname surpasses the base model IgBert from which it is derived on 8 cases. It ranks first or second on 7 of the 10 pairs. The results also show that structure is not always beneficial, with the structure-only protein language model, ProstT5, which uses the Foldseek 3Di alphabet, and the sequence-structure protein language model, SaProt, not performing as well as sequence-only models. Comparing \plmname and \almname (Table~\ref{tab:abibench_supp}), which is given all dihedral angle input rather than a summarized classification token, we also see a drop in performance. Therefore, showing the benefit of incorporating only the \name classification token into \plmname for representing proteins which differ by a few mutations. As protein language models improve with scale \citep{lin2023esm2}, it is notable that \plmname, a 420M parameter model more than 7$\times$ smaller than ESM-2 (3B), achieves better performance on average across the 10 antibody-antigens. 

\begin{table}[h]
\centering
\vspace{-2mm}
\scriptsize
\caption{Spearman correlation coefficient ($\uparrow$) for binding affinity prediction across 10 different targets from AbBiBench. The \textbf{first} and \underline{second} values are highlighted. We report the standard error across the 10 fold cross-validation in parentheses.}
\vspace{-2mm}
\label{tab:abibench}
\begin{tabular}{lcccccc>{\columncolor{gray!15}}c}
\toprule
\textbf{Target} & \textbf{ESM C} & \textbf{ESM-2 (3B)} & \textbf{SaProt} & \textbf{ProstT5} & \textbf{AbLang2} & \textbf{IgBert} & \textbf{\plmname} \\
\midrule
\textbf{1mlc} & 0.609 (0.017) & 0.551 (0.013) & 0.557 (0.020) & 0.280 (0.040) & \underline{0.634 (0.015)} & \textbf{0.665 (0.015)} & 0.616 (0.009) \\
\textbf{1n8z} & 0.673 (0.022) & 0.635 (0.019) & 0.637 (0.028) & 0.351 (0.057) & 0.646 (0.021) & \textbf{0.682 (0.023)} & \underline{0.675 (0.025)} \\
\textbf{2fxg} & \textbf{0.809 (0.010)} & 0.752 (0.010) & \underline{0.754 (0.012)} & 0.355 (0.021) & 0.752 (0.007) & 0.694 (0.013) & 0.713 (0.014) \\
\textbf{3gbn\_h1} & 0.901 (0.004) & \textbf{0.953 (0.003)} & 0.915 (0.005) & 0.638 (0.013) & 0.945 (0.004) & 0.947 (0.004) & \underline{0.948 (0.004)} \\
\textbf{3gbn\_h9} & 0.932 (0.004) & \textbf{0.971 (0.002)} & 0.952 (0.003) & 0.679 (0.017) & \underline{0.963 (0.003)} & 0.961 (0.003) & 0.962 (0.003) \\
\textbf{4fqi\_h1} & 0.871 (0.001) & \textbf{0.955 (0.001)} & 0.866 (0.001) & 0.593 (0.002) & 0.883 (0.001) & 0.898 (0.001) & \underline{0.921 (0.001)} \\
\textbf{4fqi\_h3} & 0.936 (0.001) & \textbf{0.973 (0.001)} & 0.958 (0.001) & 0.644 (0.009) & 0.969 (0.001) & 0.970 (0.001) & \underline{0.971 (0.001)} \\
\textbf{aayl49} & 0.617 (0.010) & 0.584 (0.013) & 0.584 (0.012) & 0.301 (0.014) & 0.563 (0.010) & \underline{0.611 (0.010)} & \textbf{0.625 (0.010)} \\
\textbf{aayl49\_M}L & 0.518 (0.008) & \underline{0.524 (0.008)} & 0.487 (0.009) & 0.320 (0.009) & 0.499 (0.007) & 0.524 (0.007) & \textbf{0.531 (0.007)} \\
\textbf{aayl51} & \underline{0.576 (0.007)} & 0.516 (0.009) & 0.524 (0.008) & 0.260 (0.011) & 0.527 (0.009) & 0.566 (0.010) & \textbf{0.579 (0.011)} \\
\bottomrule
\end{tabular}
\end{table}

\subsection{Controllable Sampling of Antibody Loops}

In this section, we evaluate \almname on its ability to guide the structure of the loop at the residue level, by analyzing if sampled loops are consistent in structure to the masked out loop.

\xhdr{Experiment setup} For the CDR1, CDR2, and CDR3 of the heavy and light chain, we randomly sample 50 structures from SAbDab, which are in the test set of \name. \name $\mathbf{t}, \mathbf{X}$ tokens of the dihedral angles and masked sequence are inputted to \almname. We sample loop sequences from the resulting amino acid likelihoods of \almname for each of these antibodies and fold the sampled sequence with Ibex.

\begin{figure}[H]
  \centering 
    \vspace{-2mm}
  \includegraphics[width=0.65\textwidth]{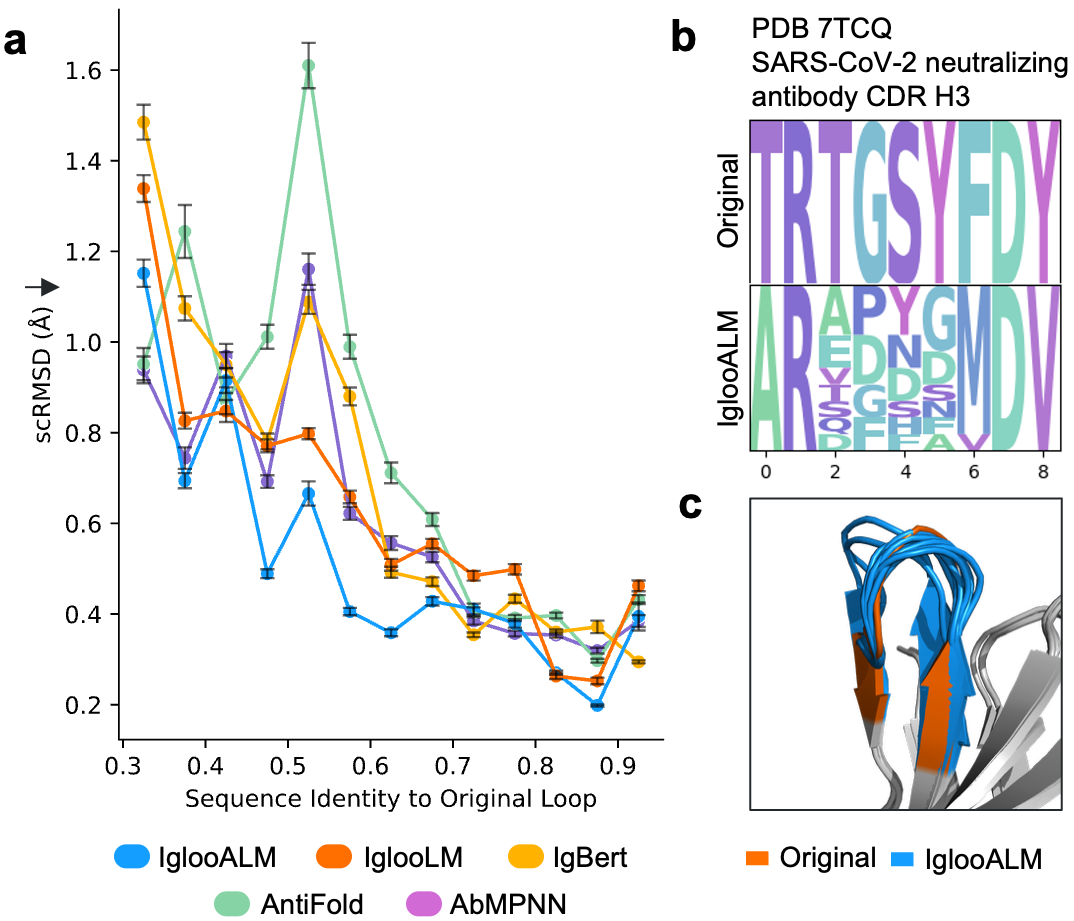}
  \caption{
  \textbf{a} Mean self-consistency (sc) RMSD (\AA) $\downarrow$ of sampled loop sequences compared to original loop structures across sequence identity bins. Error bars show standard error of the mean across the generated structures aggregated in each sequence identity bin.
  \textbf{b} Sequence logo of original and ten \almname sampled sequences of the CDR H3 loop region for a SARS-CoV-2 neutralizing antibody (PDB 7TCQ) at $\lambda=0.5$.
  \textbf{c} Predicted structure of the CDR H3 loop regions aligned to PDB 7TCQ.}
  \label{fig:loop_rmsd}
  \vspace{-5mm}
\end{figure}

\xhdr{Results}
We show the scRMSD of the sampled loops stratified by sequence identity (Fig.~\ref{fig:loop_rmsd}a). \almname excels at generating loops at different levels of sequence diversity while maintaining a similar structure, improving both on state-of-the-art antibody inverse folding models and the base model. We highlight a particular example for the redesign of the H3 loop of a SARS-CoV-2 neutralizing antibody from the PDB structure 7TCQ. At a sampling temperature of 0.5, \almname samples loops with an average edit distance of 6.6 from the loop of length 9 (Fig.~\ref{fig:loop_rmsd}b). The predicted structures of the sampled loops maintain the beta hairpin structure of the original loop with an average loop RMSD of 0.79 \AA\ (Fig.~\ref{fig:loop_rmsd}c). Other examples of sampled loops and their structures are presented in Fig.~\ref{fig:loop_rmsd_supp}.

\xhdr{Baselines} We evaluate against recently published state-of-the-art inverse folding models for antibodies: AbMPNN \citep{abmpnn}--a version of ProteinMPNN \citep{proteinmpnn} finetuned on antibody structures, and AntiFold \citep{hoie2024antifold}--a version of ESM-IF1 \citep{esmif1} finetuned on antibody structures. For these models, we only generate the loop sequence given the full backbone and the sequence of the rest of the antibody. We also compare to the base model, IgBert, and \plmname which does not include the multimodal residue tokens $\mathbf{x}$. 

\xhdr{Evaluation setup} For each loop and model, we sample 10 sequences at the following sampling temperatures: $\lambda=0.01, 0.05, 0.1, 0.2, 0.5, 1.0, 2.0$. In total, for each model, we generate $50\times 10 \times 6 \times 7 = 21,000$ sequences for the different structures, sequence samples, loop types, and temperatures, respectively. Then we align the generated loop regions with the original structure and evaluate the self-consistency (sc) RMSD between the two loop structures. Sampling sequences at different temperatures is necessary to generate sequences with different levels of sequence identity to the original loop, since recapitulating the original loop sequence would achieve low scRMSD but would not be useful for the design of new H3 loops.

\section{Conclusion}
Here we present \name, a multimodal tokenizer for antibody loops with a novel contrastive learning objective based on dihedral angle distance between loop backbones. Applying \name, we achieve state-of-the-art results in retrieving similar loop conformations and recover known canonical clusters. \name tokens can also be incorporated into protein language models for improved binding affinity predictions with \plmname and for controllable generation of antibody loops with \almname. While \almname demonstrates strong in silico results, more comprehensive wet-lab validation evaluation is needed to evaluate whether redesigned antibodies maintain binding with antigens. \name could be further extended to incorporate other modalities such as all-atom structure, epitope information, function, and binding affinity. We also envision \name could be applied to lead optimization and affinity maturation of antibodies \citep{makowski2022co,dreyer2025computational,frey2025lab}. By introducing multimodal loop tokens, \name opens new directions for multimodal foundation models for rational antibody design. 

\subsection*{Acknowledgments}
We are grateful to Joshua Southern, Sidney Lisanza, Nathan Frey, Saeed Saremi, Sarah Robinson, Imee Sinha, Catherine Wong, Pan Kessel, Leon Hetzel, Franziska Seeger, and the rest of the Prescient Design team for useful discussions.

\subsection*{Reproducibility Statement}
All code for data processing, implementation, training scripts, evaluation and analyses scripts, and for reproducing results in the paper is available at \url{https://github.com/prescient-design/igloo}. Details of the dataset used for training \name are provided in Appendix \ref{appendix:data-igloo}, and for training \plmname and \almname are provided in Appendix \ref{appendix:data-plm}. Details for processing of AbBiBench data are available at Appendix \ref{appendix:abbibench-data}. Training details, hyperparameters, GPUs used, and training duration for training \name, \plmname, \almname, and regression models for AbBiBench are available at Appendix \ref{appendix:implementation}. The transformer architecture used in \name is based on the \texttt{TransformerLayer} from ESM-2 available at \url{https://github.com/facebookresearch/esm}. The base model, IgBert \citep{kenlay2024igbert}, is publicly available at \url{https://huggingface.co/Exscientia/IgBert}. Code for finetuning IgBert for \plmname and \almname is available on our GitHub. Full details of the \name loss function and objective are available at Section~\ref{sec:igloo-objective}. All evaluation metrics used for experiments are specified explicitly in Section~\ref{sec:experiments}. 

\subsection*{Ethics Statement}
\name is a method for tokenizing loop regions of antibodies and may be used for antibody design. All antibody sequences used to develop and evaluate \name were obtained from publicly available databases and contain no personal or patient‐identifiable information. No new animal or human subjects were involved. Methods that facilitate antibody engineering can present dual-use concerns. Here we present use cases where \name is applied for achieving positive impact.

\bibliography{iclr2025_conference}
\bibliographystyle{iclr2025_conference}

\appendix
\renewcommand{\thetable}{S\arabic{table}}
\setcounter{table}{0}
\renewcommand{\thefigure}{S\arabic{figure}}
\setcounter{figure}{0}

\section{Dataset Processing}
\subsection{\name Training Data} \label{appendix:data-igloo}
We process 18,303 structures from SAbDab and STCRDab, which are comprised of 14,341 antibodies, 3,095 nanobodies, and 867 TCRs. From these structures, we run ANARCI~\citep{dunbar2016anarci} on the sequences to identify the loop regions (CDR1, CDR2, CDR3, CDR4) in the North definition~\cite{north2011new} from their AHo alignment~\citep{honegger2001yet}. A valid loop requires defined $\phi,\psi,\omega$ angles and at least 5 residues before and after the loop, referred to as the stem region, yielding 108,167 loop structures. To define a train, test, and validation split, we cluster on the concatenated CDR sequences with MMseqs2~\citep{steinegger2017mmseqs2}, using an 80\% sequence identity threshold.

Ibex~\citep{ibex} predicted CDR loops from antibodies in paired OAS are also included in the \name training set. To maximize sequence diversity of the predicted loop structures, they are downsampled from an initial set of 2,447,258 down to 87,456 by clustering on both concatenated CDR sequences as well as H3 loop sequences with MMseqs2~\citep{steinegger2017mmseqs2}, using a 50\% sequence identity threshold. In total, we include 699,648 predicted loop structures in the training set.

\begin{table}[h]
\centering
\caption{Number of each loop type in the \name training dataset from SAbDab, STCRDab and paired OAS.}
\vspace{1mm}
\label{tab:loop_counts}
\begin{tabular}{lcc}
\toprule
\textbf{Loop Type} & \textbf{SAbDab and STCRDab} & \textbf{Paired OAS} \\
\midrule
H1 & 14,877 & 87,456 \\
H2 & 14,876 & 87,456 \\
H3 & 14,875 & 87,456 \\
H4 & 14,877 & 87,456 \\
L1 & 12,167 & 87,456 \\
L2 & 12,168 & 87,456 \\
L3 & 12,159 & 87,456 \\
L4 & 12,168 & 87,456 \\
\midrule
Total & 108,167 & 699,648 \\
\bottomrule
\vspace{-4mm}
\end{tabular}
\end{table}

\subsection{\plmname and \almname Training Data} \label{appendix:data-plm}
We fold the heavy and light chain with Ibex for 2,447,258 antibodies. The heavy chains and light chains are clustered separately with MMseqs2~\citep{steinegger2017mmseqs2}, using a 90\% sequence identity threshold. This results in 247,156 light chain clusters and 875,767 heavy chain clusters. We randomly sample 10,000 light chain and 20,000 heavy chain clusters for the validation and test sets, respectively. For the training set, we keep all sequences in the sequence identity clusters, and for the validation and test sets, we only keep the representative sequence from each cluster. In total, we train \plmname and \almname on 4,598,332 antibody chains.

\subsection{AbBiBench Data} \label{appendix:abbibench-data}
We use AbBiBench \citep{abibench} benchmark, which has for an antibody-antigen pair, heavy chain mutant sequences and their binding affinity score. The binding affinity score is the $-\log K_d$ for all antibody-antigen pairs except for 2fjg and 1mlc, which is $\log$ enrichment. For some antibody-antigens, we filter out sequences that do not have binding affinity scores and are given default scores instead. The final number of sequences and filtered used for each antibody-antigen target is shown in Table~\ref{tab:abibench-counts}. We train models for 10 out of 11 antibody-antigens in AbBiBench. The Integrin-$\alpha$-1 AQC2 antibody-antigen dataset is not tested due to an insufficient number of binding affinity measurements ($N=40$). For each sequence, we fold the heavy chain with the light chain from the structure in the PDB ID with Ibex and extract the structures of the loops for \plmname. All binding affinity values for an antibody-antigen pair are scaled by subtracting the mean of the training distribution and scaling to unit variance.
\begin{table}[h]
\centering
\caption{Number of sequences for each antibody-antigen in the AbBiBench benchmarking dataset.} \label{tab:abibench-counts}
\scriptsize
\vspace{1mm}
\label{tab:abibench_counts}
\begin{tabular}{lllcc}
\toprule
\textbf{PDB ID} & \textbf{Seed antibody} & \textbf{Antigen} & \textbf{Number of sequences} & \textbf{Filtered out values}\\
\midrule
1n8z & Trastuzumab & HER2 & 419 & -\\
1mlc & D44.1 & Hen-egg-white lysozyme & 1,229 & - \\
2fjg & G6.31 & VEGF & 2,223 & - \\
3gbn\_h1 & CR6261 & Influenza A/New Caledonia/20/99 (H1N1) & 1,673 & 7.0 \\
3gbn\_h9 & CR6261 & Influenza A/Hong Kong/1073/1999 (H9N2) & 1,470 & 7.0 \\
4fqi\_h1 & CR9114 & Influenza A/New Caledonia/20/99 (H1N1) & 63,419 & 7.0 \\
4fqi\_h3 & CR9114 & Influenza A/Wisconsin/67/2005 (H3N2) & 7,174 & 6.0 \\
aayl49 & AAYL49 & Spike HR2 & 4,312 & - \\
aayl49\_ML & AAYL49\_ML & Spike HR2 & 8,953 & - \\
aayl51 & AAYL51 & Spike HR2 & 4,320 & - \\
\bottomrule
\vspace{-4mm}
\end{tabular}
\end{table}

\section{Implementation Details} \label{appendix:implementation}
\subsection{Training \name}
\name is trained for 100 epochs on 1 NVIDIA H100. We set the following hyperparameters for training \name to be dihedral temperature (0.1), unit circle regularization weight (0.01), number of transformer layers (4), codebook commit loss weight (0.5), max loop length (36), and batch size (64). The following hyperparameters were chosen from: learning rate  ($10^{-5}-10^{-3}$), embedding dimension (32, 128, 1024), codebook size (1024, 8192), and weight decay ($0,10^{-5}$). We select the checkpoint at the epoch with the lowest validation loss and select the best hyperparameter based on the average recovery of the canonical clusters \citep{kelow2022penultimate} on the validation set. We use a two-phase training approach; in the first phase, the model is trained on the SAbDab and paired OAS dataset, and in the second phase, the model is trained only on SAbDab. We use an embedding dimension of 128 and a codebook size of 8192. In the first stage, the learning rate is $5\times 10^{-5}$ and weight decay $0$, and for the second stage, a learning rate of $5\times 10^{-5}$ and weight decay of $10^{-5}$ is used.

\subsection{Training \plmname and \almname}
We take the publicly available pretrained weights and hyperparameters from IgBert \citep{kenlay2024igbert} and continue to finetune the model with the \name tokens for \plmname, and with \name tokens and multimodal residue tokens for \almname. Both models were trained for 3 days on 4 NVIDIA H100s, which correspond to 53k steps over 5 epochs.

\subsection{Paratope Retrieval} \label{appendix:retrieval-method}
For obtaining embeddings from the baseline models ESM C, ESM-2 (3B), AbLang2, IgBert, ProstT5, AbLang2, and IgBert, we use publicly available weights and embed the whole antibody chain and use the mean embedding of the loop residues as the loop embedding. For the structure tokenizers MIF, ProteinMPNN, and Amino Aseed (continuous tokens) we use the implementation provided by \citep{yuan2025protein} and average residue-level tokens over the loop region to obtain a loop embedding. For Foldseek 3Di, since the embeddings are only 2-dimensional, we concatenate the flattened representations of all of the residues in the loop region to obtain the loop embedding.

\subsection{Training Regression Models for AbBiBench} \label{appendix:abbibench-method}
For the embeddings of each model and antibody-antigen target, we train a ridge regression with a 10-fold nested cross-validation. The 10 outer folds are used for testing, each containing a 5-fold inner cross-validation that selects the optimal L2 penalty $\lambda\in\{1, 10^{-1}, 10^{-2}, \dots, 10^{-6}, 0\}$ within that fold. For every outer fold, the model was retrained with its fold-specific best $\lambda$ on the entire training partition, scored on the held-out test partition, and the Spearman correlation coefficient, $\rho$, is averaged across the 10 folds.

\section{Ablation Study} \label{sec:ablation}
To evaluate the contributions of the components of \name, we conduct the following ablation studies to understand the effect of (1) the dihedral distance contrastive loss, (2) distance-threshold filtering of positive ($\mathcal{D}<0.1$) and negative ($\mathcal{D}>0.47$) pairs, (3) the sequence modality track, (4) the dihedral angle modality track, and (5) only defining positive pairs between loops of the same length (Appendix~\ref{sec:loop-length-independent}). We evaluate the ablated models with the same experimental setup as outlined in Section~\ref{exp:paratope}.

\begin{table}[H]
\centering
\caption{Average precision at rank 1, 5, 10, and 20 for retrieval of similar CDR paratopes evaluated with $\text{RMSD}<1\text{\AA}$ and $\mathcal{D} < 0.47$. The \textbf{first} and \underline{second} best performance are highlighted below. CL is contrative learning, DT filter is distance-threshold filter of positive ($\mathcal{D}<0.1$) and negative ($\mathcal{D}>0.47$) pairs, and loop length refers to training \name with positive pairs defined between loops of different lengths (Appendix~\ref{sec:loop-length-independent}).}
\vspace{1mm}
\label{tab:ablation}
\tiny
\vspace{1mm}
\label{tab:retrieval-top10}
\begin{tabular}{ll|cccccc|cccccc}
\toprule
 & & \multicolumn{6}{c|}{\textbf{\% RMSD $<$ 1~\AA}} & \multicolumn{6}{c}{$\%\,\mathbf{\mathcal{D}<0.47}$} \\
 \multicolumn{2}{l|}{\textbf{\# Loops retrieved}} & \textbf{L1} & \textbf{L2} & \textbf{L3} & \textbf{H1} & \textbf{H2} & \textbf{H3} & \textbf{L1} & \textbf{L2} & \textbf{L3} & \textbf{H1} & \textbf{H2} & \textbf{H3} \\
\midrule
\multirow{7}{*}{\textbf{1}} & \textbf{Random} &
0.518 & 0.603 & 0.417 & 0.221 & 0.321 & 0.152 &
0.669 & 0.770 & 0.459 & 0.597 & 0.435 & 0.157 \\

& \textbf{No CL loss} &
0.866 & 0.732 & \textit{0.714} & 0.418 & \underline{0.697} & 0.305 &
0.915 & \textit{0.967} & \textit{0.836} & 0.686 & \textit{0.898} & 0.464 \\

& \textbf{No DT filter} &
\underline{0.877} & \textbf{0.748} & 0.673 & 0.452 & 0.680 & \underline{0.330} &
0.919 & \textbf{0.994} & 0.789 & \underline{0.862} & 0.882 & \underline{0.539} \\

& \textbf{Sequence only} &
0.790 & \underline{0.740} & 0.666 & 0.413 & 0.472 & 0.219 &
0.809 & 0.956 & 0.719 & 0.750 & 0.557 & 0.292 \\

& \textbf{Dihedral angles only} &
0.870 & 0.702 & 0.591 & \textit{0.557} & 0.590 & 0.298 &
\textbf{0.945} & 0.936 & 0.752 & \textit{0.841} & 0.741 & 0.491 \\

& \textbf{Mismatched length} &
\textbf{0.878} & \textit{0.734} & \underline{0.749} & \underline{0.593} & \textbf{0.701} & \textbf{0.339} &
\underline{0.941} & \underline{0.993} & \textbf{0.870} & 0.834 & \textbf{0.928} & \textit{0.523} \\

\rowcolor{gray!15} & \textbf{\name} &
\textit{0.871} & \textbf{0.748} & \textbf{0.761} & \textbf{0.603} & \textit{0.691} & \textit{0.327} &
\textit{0.935} & \textbf{0.993} & \underline{0.856} & \textbf{0.885} & \underline{0.918} & \textbf{0.669} \\
\midrule
\multirow{6}{*}{\textbf{5}} & \textbf{Random} &
0.558 & 0.556 & 0.387 & 0.229 & 0.347 & 0.138 &
0.679 & 0.742 & 0.401 & 0.561 & 0.480 & 0.136 \\

& \textbf{No CL loss} &
0.826 & 0.724 & \textit{0.636} & 0.420 & 0.611 & 0.279 &
\textit{0.897} & \textit{0.973} & 0.765 & 0.677 & \underline{0.881} & 0.413 \\

& \textbf{No DT filter} &
\textit{0.829} & \textbf{0.748} & 0.629 & 0.453 & \underline{0.646} & \underline{0.315} &
\underline{0.904} & \textbf{0.996} & \textit{0.784} & \underline{0.788} & \textit{0.876} & \underline{0.506} \\

& \textbf{Sequence only} &
0.798 & 0.704 & 0.569 & 0.401 & 0.477 & 0.203 &
0.844 & 0.909 & 0.663 & 0.633 & 0.573 & 0.245 \\

& \textbf{Dihedral angles only} &
\underline{0.837} & 0.682 & 0.611 & \textit{0.463} & 0.565 & \textit{0.280} &
0.894 & 0.917 & 0.759 & \textit{0.786} & 0.728 & 0.454 \\

& \textbf{Mismatched length} &
0.828 & \textit{0.740} & \underline{0.646} & \underline{0.473} & \textit{0.627} & \textbf{0.316} &
\textit{0.897} & \underline{0.993} & \textbf{0.828} & 0.643 & 0.833 & \textit{0.495} \\

\rowcolor{gray!15} & \textbf{\name} &
\textbf{0.841} & \underline{0.743} & \textbf{0.666} & \textbf{0.501} & \textbf{0.658} & \underline{0.315} &
\textbf{0.909} & \underline{0.993} & \underline{0.827} & \textbf{0.805} & \textbf{0.923} & \textbf{0.553} \\
\midrule
\multirow{6}{*}{\textbf{10}} & \textbf{Random} &
0.550 & 0.556 & 0.384 & 0.235 & 0.345 & 0.132 &
0.666 & 0.726 & 0.399 & 0.556 & 0.499 & 0.133 \\

& \textbf{No CL loss} &
0.796 & 0.723 & \textit{0.601} & 0.424 & 0.559 & 0.265 &
0.866 & 0.975 & \textit{0.724} & 0.672 & \textit{0.795} & 0.376 \\

& \textbf{No DT filter} &
0.802 & \textbf{0.744} & 0.584 & \textit{0.460} & \underline{0.600} & \underline{0.302} &
\textbf{0.884} & \textbf{0.994} & 0.717 & \textit{0.728} & \underline{0.840} & \underline{0.470} \\

& \textbf{Sequence only} &
0.781 & 0.705 & 0.503 & 0.391 & 0.494 & 0.198 &
0.852 & 0.898 & 0.571 & 0.655 & 0.613 & 0.239 \\

& \textbf{Dihedral angles only} &
\textbf{0.812} & 0.672 & 0.595 & 0.442 & 0.541 & 0.263 &
\underline{0.880} & 0.905 & 0.715 & \textbf{0.757} & 0.710 & 0.400 \\

& \textbf{Mismatched length} &
\textit{0.804} & \textit{0.737} & \underline{0.604} & \underline{0.469} & \textit{0.564} & \textbf{0.304} &
\underline{0.867} & \underline{0.992} & \underline{0.735} & 0.671 & 0.759 & \textit{0.462} \\

\rowcolor{gray!15} & \textbf{\name} &
\underline{0.809} & \underline{0.742} & \textbf{0.623} & \textbf{0.473} & \textbf{0.620} & \textit{0.300} &
\textit{0.879} & \underline{0.993} & \textbf{0.764} & \underline{0.736} & \textbf{0.854} & \textbf{0.473} \\
\midrule
\multirow{7}{*}{\textbf{20}} & \textbf{Random} &
0.545 & 0.557 & 0.373 & 0.249 & 0.351 & 0.127 &
0.648 & 0.730 & 0.392 & 0.559 & 0.508 & 0.126 \\

& \textbf{No CL loss} &
0.780 & 0.689 & 0.516 & 0.410 & 0.496 & 0.242 &
0.845 & 0.927 & 0.603 & 0.649 & 0.700 & 0.335 \\

& \textbf{No DT filter} &
0.788 & \underline{0.704} & \underline{0.543} & \underline{0.459} & \underline{0.562} & \underline{0.279} &
\textbf{0.860} & \textit{0.950} & \underline{0.635} & 0.686 & \textit{0.747} & \textbf{0.417} \\

& \textbf{Sequence only} &
0.761 & 0.693 & 0.484 & 0.400 & 0.482 & 0.193 &
0.828 & 0.887 & 0.533 & 0.614 & 0.624 & 0.217 \\

& \textbf{Dihedral angles only} &
\textbf{0.795} & 0.651 & 0.524 & 0.424 & 0.497 & 0.245 &
\underline{0.853} & 0.884 & 0.617 & \underline{0.702} & 0.671 & 0.356 \\

& \textbf{Mismatched length} &
\textit{0.789} & \textit{0.703} & \textit{0.533} & \textbf{0.465} & \textit{0.538} & \textbf{0.280} &
0.843 & \underline{0.954} & \textit{0.627} & \textit{0.689} & \underline{0.749} & \underline{0.408} \\

\rowcolor{gray!15} & \textbf{\name} &
\underline{0.793} & \textbf{0.705} & \textbf{0.558} & \underline{0.459} & \textbf{0.578} & \textit{0.278} &
\textit{0.851} & \textbf{0.956} & \textbf{0.674} & \textbf{0.715} & \textbf{0.749} & \textit{0.402} \\
\bottomrule
\end{tabular}
\vspace{+1mm}
\end{table}

\xhdr{Dihedral distance contrastive loss} Key to the \name approach is the contrastive learning objective for the model to learn to place loops that share similar backbone dihedral angles in the same region of the latent space. To test this component of the model, we removed the dihedral contrastive loss from training. Consequently, the ablated model is only focused on the reconstruction of masked amino acids and dihedral angles. We observed in Table~\ref{tab:ablation} that the contrastive learning objective improves performance across loop regions on both precision for $\text{RMSD}<1\text{\AA}$ and $\mathcal{D} < 0.47$, with improvements of $11.8\%$ on the L3 loop and $20.0\%$ on the H3 loop in precision at rank 20.

\xhdr{Distance-threshold filtering of positive ($\mathcal{D}<0.1$) and negative ($\mathcal{D}>0.47$) pairs}
$\mathcal{D}$ is a continuous measure of the difference in dihedral angles between two loop backbones. In order for the model to not overfit to an arbitrary threshold of 0.47, which was established by \cite{kelow2022penultimate}, we established a distance-threshold filter where loops with $0.1\leq\mathcal{D}\leq0.47$ are ignored. We train an ablated model where positive ($\mathcal{D}\leq 0.47$) and negative ($\mathcal{D}>0.47$) pairs and find that performance is generally comparable to when distance-thresholding is applied, with \name offering slight improvements across most loop types.

\xhdr{Multimodal learning in \name}
In \name the input to the transformer is $\mathbf{X} =\mathbf{D} + \mathbf{A}$, in this section we remove the dihedral angles, $\mathbf{D}$, and sequence, $\mathbf{A}$, from the model separately. We also adjust the loss function correspondingly. The masked reconstruction of dihedral angles and the masked reconstruction of amino acid identities objectives are also removed, respectively. In Table~\ref{tab:ablation}, we observe the dihedral angle modality is most important to the retrieval task, notably for the H3 loop retrieval with an improvement of 85.2\%. The addition of the sequence modality is also helpful with improvements observed for almost all loop types and on H3 loop retrieval, an improvement of 12.7\% is observed.

\section{\name with Mismatched Loop Length} \label{sec:loop-length-independent}
The \name contrastive loss function only assigns positive labels to pairs of loops of the same length. However, \cite{nowak2016length} explore CDR clusters with loops of multiple lengths, and find clusters L1–10,11,12-A; L1–13,14-A; L3–9,10-A; and L3–10,11-A with loops of different lengths. In this section, we show how \name can be trained to align loops of different lengths in the latent space.

To define positive pairs for loops of different lengths, we use the approach from \cite{nowak2016length}. Loops are aligned by the $\mathrm{C}_\alpha$ coordinates of their stem region, which we define as the $N_{\text{stem}}$ amino acids before and after the loop. We then use dynamic time warping (DTW) \citep{giorgino2009computing} to determine an alignment between the loop $\mathrm{C}_\alpha$ coordinates. For the aligned residues, the dihedral distance $\mathcal{D}$ is calculated (Algorithm~\ref{alg:alignloops}). Finally, positive and negative pairs are defined by thresholds on $\mathcal{D}$.

\begin{algorithm}[t]
\caption{Dihedral angle distance between a pair of loops of different lengths}
\label{alg:alignloops}
\textbf{Input:} loop dihedral angles $\boldsymbol\phi_1,  \boldsymbol\psi_1, \boldsymbol\omega_1 \in (-\pi, \pi]^{n},\; \boldsymbol\psi_2, \boldsymbol\phi_2, \boldsymbol\omega_2 \in (-\pi, \pi]^{m}$; loop $\mathrm{C}_\alpha$ coordinates $\mathbf{L}_1\!\in\!\mathbb{R}^{n\times3}$, $\mathbf{L}_2\!\in\!\mathbb{R}^{m\times3}$; stem $\mathrm{C}_\alpha$ coordinates $\mathbf{S}_1,\mathbf{S}_2\!\in\!\mathbb{R}^{N_{\mathrm{stem}}\times3}$; tolerance $k$ (max. residue length difference)  \\
\textbf{Output:} dihedral angle distance $\mathcal{D} \in [0,4]$
\begin{algorithmic}[1]
\If{$|n-m| > k$} \State \Return \texttt{4.0}  \Comment{returns $\mathcal{D}_\text{max}$}  \EndIf
\State $\boldsymbol\mu_1 \gets \mathrm{mean}(\mathbf{S}_1)$; \quad $\boldsymbol\mu_2 \gets \mathrm{mean}(\mathbf{S}_2)$
\State $\tilde{\mathbf{S}}_1 \gets \mathbf{S}_1 - \boldsymbol\mu_1$; \quad $\tilde{\mathbf{S}}_2 \gets \mathbf{S}_2 - \boldsymbol\mu_2$
\State $(\mathbf{R},\mathbf{t},\text{RMSD}_\text{stem}) \gets \mathrm{Kabsch}(\tilde{\mathbf{S}}_1,\tilde{\mathbf{S}}_2)$
\If{\text{RMSD} $> 1.0$ \AA} \State \Return \texttt{4.0} \Comment{returns $\mathcal{D}_\text{max}$} \EndIf
\State $\tilde{\mathbf{L}}_1 \gets (\mathbf{L}_1-\boldsymbol\mu_1)\mathbf{R}^\top + \boldsymbol\mu_2 + \mathbf{t}$; \quad
       $\tilde{\mathbf{L}}_2 \gets \mathbf{L}_2$
\State $\mathcal{P} \gets \mathrm{DTW}(\tilde{\mathbf{L}}_1,\tilde{\mathbf{L}}_2)$ \Comment{warping path $\mathcal{P}$ mapping residues of $\mathbf{L}_1$ to $\mathbf{L}_2$}
\State $\tilde{\boldsymbol{\phi}}_1 \gets \mathcal{P}(\boldsymbol{\phi}_1); \quad \tilde{\boldsymbol{\psi}}_1 \gets \mathcal{P}(\boldsymbol{\psi}_1);  \quad \tilde{\boldsymbol{\omega}}_1 \gets \mathcal{P}(\boldsymbol{\omega}_1);$
\State $\mathcal{D}_\phi \gets  \mathrm{mean}\bigl(2\bigl(1 - \cos(\tilde{\boldsymbol{\phi}}_1 - {\boldsymbol{\phi}}_2 )\bigr)\bigr)$; \quad $\mathcal{D}_\psi \gets \mathrm{mean}\bigl(2\bigl(1 - \cos(\tilde{\boldsymbol{\psi}}_1 - {\boldsymbol{\psi}}_2 )\bigr)\bigr)$; \quad $\mathcal{D}_\omega \gets \mathrm{mean}\bigl(2\bigl(1 - \cos(\tilde{\boldsymbol{\omega}}_1 - {\boldsymbol{\omega}}_2 )\bigr)\bigr)$
\State $\mathcal{D} \gets \mathrm{mean}(\mathcal{D}_\phi, \mathcal{D}_\psi, \mathcal{D}_\omega)$
\State \Return $\mathcal{D}$
\end{algorithmic}
\end{algorithm}

We train the \name architecture with $N_{\text{stem}}=5$ and tolerance $k=1$, which is consistent with the multi-length clusters found by \cite{nowak2016length}. For batches in an epoch, we find on average 5.0\% of pairs of loops in the batch to be of different lengths and $\mathcal{D} < 0.1$. In Table~\ref{tab:ablation} we observe that training \name with positive pairs defined between loops of different lengths leads to a slight decay in performance on most loop types.

\section{Additional Results}

\subsection{Paratope Retrieval}
We present additional results for the retrieval of similar loops evaluated with precision at rank 1, 5, and 10 (Table~\ref{tab:retrieval-topk}).

\begin{table}[h]
\centering
\caption{Average precision at rank 1, 5, and 10 for retrieval of similar CDR paratopes.  
Models are shown in rows.  The \textbf{first}, \underline{second}, and \textit{third} best performance for each column are highlighted.}
\tiny
\vspace{-3mm}
\label{tab:retrieval-topk}
\begin{tabular}{ll|cccccc|cccccc}
\toprule
 & & \multicolumn{6}{c|}{\textbf{\% RMSD $<$ 1~\AA}} & \multicolumn{6}{c}{$\%\,\mathbf{\mathcal{D}<0.47}$} \\
 \multicolumn{2}{l|}{\textbf{\# Loops retrieved}} & \textbf{L1} & \textbf{L2} & \textbf{L3} & \textbf{H1} & \textbf{H2} & \textbf{H3} & \textbf{L1} & \textbf{L2} & \textbf{L3} & \textbf{H1} & \textbf{H2} & \textbf{H3} \\
\bottomrule
\multirow{11}{*}{\textbf{1}} & \textbf{ESM C}          & 0.765 & 0.747 & 0.561 & 0.420 & 0.670 & 0.242 & 0.860 & 0.969 & 0.533 & 0.549 & 0.786 & 0.254 \\
& \textbf{ESM-2 (3B)}     & 0.706 & 0.733 & 0.525 & 0.420 & 0.586 & 0.238 & 0.793 & 0.960 & 0.505 & 0.566 & 0.867 & 0.228 \\
& \textbf{AbLang2}        & 0.766 & 0.747 & 0.535 & 0.384 & 0.567 & 0.191 & 0.835 & 0.919 & 0.533 & 0.575 & 0.825 & 0.212 \\
& \textbf{IgBert}         & 0.802 & 0.733 & 0.621 & 0.410 & 0.428 & 0.194 & 0.853 & 0.971 & 0.597 & 0.528 & 0.653 & 0.252 \\
& \textbf{SaProt}         & 0.804 & 0.747 & 0.696 & 0.475 & 0.525 & 0.271 & 0.877 & 0.969 & 0.722 & 0.733 & 0.674 & 0.341 \\
& \textbf{ProstT5}        & 0.791 & \textbf{0.748} & 0.729 & \textit{0.531} & \underline{0.703} & 0.331 & \textit{0.920} & 0.987 & \textbf{0.878} & 0.748 & 0.871 & 0.487 \\
& \textbf{MIF}            & \underline{0.879} & 0.747 & 0.643 & \underline{0.557} & 0.564 & 0.290 & 0.915 & 0.991 & 0.806 & \underline{0.911} & 0.837 & 0.381 \\
& \textbf{ProteinMPNN}    & 0.863 & 0.733 & 0.706 & 0.468 & 0.693 & \textit{0.355} & 0.906 & 0.958 & 0.798 & \textbf{0.915} & \textit{0.900} & \textit{0.543} \\
& \textbf{Foldseek 3Di}  & 0.840 & \underline{0.748} & \underline{0.739} & 0.509 & \textbf{0.713} & \underline{0.361} & \textbf{0.947} & \textbf{0.994} & \underline{0.859} & 0.771 & 0.874 & \underline{0.585} \\
& \textbf{Amino Aseed}   & \textbf{0.890} & 0.734 & \textit{0.738} & 0.462 & \textit{0.703} & \textbf{0.374} & 0.907 & \textbf{0.994} & 0.824 & 0.787 & \underline{0.910} & 0.528 \\
\rowcolor{gray!15} & \textbf{\name}        & \textit{0.871} & \textit{0.748} & \textbf{0.761} & \textbf{0.603} & 0.691 & 0.327 & \underline{0.935} & \underline{0.993} & \textit{0.856} & \textit{0.885} & \textbf{0.918} & \textbf{0.669} \\
\midrule
\multirow{11}{*}{\textbf{5}} & 0.765 & 0.710 & 0.610 & 0.408 & 0.627 & 0.214 & 0.842 & 0.900 & 0.619 & 0.658 & 0.748 & 0.256 \\
 & \textbf{ESM-2 (3B)}  & 0.730 & 0.711 & 0.540 & 0.447 & 0.562 & 0.225 & 0.837 & 0.910 & 0.607 & 0.697 & 0.788 & 0.251 \\
& \textbf{AbLang2}        & 0.720 & 0.717 & 0.588 & 0.402 & 0.563 & 0.188 & 0.786 & 0.896 & 0.595 & 0.604 & 0.816 & 0.238 \\
& \textbf{IgBert}         & 0.776 & 0.709 & 0.570 & 0.397 & 0.578 & 0.182 & 0.845 & 0.938 & 0.566 & 0.677 & 0.798 & 0.259 \\
& \textbf{SaProt}         & 0.782 & 0.741 & 0.607 & 0.441 & 0.552 & 0.246 & 0.858 & 0.970 & 0.654 & 0.698 & 0.753 & 0.316 \\
& \textbf{ProstT5}        & 0.805 & \textbf{0.745} & 0.643 & \textit{0.499} & \underline{0.671} & 0.302 & 0.879 & \textit{0.985} & 0.772 & 0.742 & \textit{0.841} & 0.445 \\
& \textbf{MIF}            & 0.825 & 0.739 & 0.652 & 0.470 & 0.571 & 0.263 & \underline{0.886} & 0.985 & 0.787 & 0.755 & 0.750 & 0.361 \\
& \textbf{ProteinMPNN}    & \underline{0.851} & 0.742 & 0.651 & 0.482 & 0.648 & \textit{0.327} & 0.878 & 0.974 & 0.787 & \textit{0.762} & 0.836 & \textit{0.472} \\
& \textbf{Foldseek 3Di}       & 0.810 & 0.742 & \textbf{0.666} & \textbf{0.530} & \textbf{0.680} & \underline{0.326} & \underline{0.886} & 0.985 & \underline{0.815} & \underline{0.784} & \underline{0.869} & \underline{0.475} \\
& \textbf{Amino Aseed}    & \textbf{0.852} & \underline{0.745} & \underline{0.657} & 0.469 & 0.606 & \textbf{0.336} & 0.870 & \textbf{0.994} & \textit{0.813} & 0.736 & 0.825 & 0.469 \\
\rowcolor{gray!15} & \textbf{\name} & \textit{0.841} & \textit{0.743} & \textbf{0.666} & \underline{0.501} & \textit{0.658} & 0.315 & \textbf{0.909} & \underline{0.993} & \textbf{0.827} & \textbf{0.805} & \textbf{0.923} & \textbf{0.553} \\
\midrule 
\multirow{11}{*}{\textbf{10}} & 0.768 & 0.705 & 0.537 & 0.418 & 0.583 & 0.202 & 0.828 & 0.921 & 0.572 & 0.678 & 0.768 & 0.230 \\
& \textbf{ESM-2 (3B)} & 0.744 & 0.713 & 0.521 & 0.439 & 0.561 & 0.210 & 0.837 & 0.895 & 0.569 & 0.716 & 0.734 & 0.237 \\
& \textbf{AbLang2}  & 0.719 & 0.686 & 0.522 & 0.408 & 0.553 & 0.186 & 0.799 & 0.844 & 0.558 & 0.611 & 0.764 & 0.228 \\
& \textbf{IgBert}   & 0.727 & 0.691 & 0.520 & 0.416 & 0.542 & 0.182 & 0.806 & 0.920 & 0.541 & 0.708 & 0.726 & 0.245 \\
& \textbf{SaProt}   & 0.749 & 0.733 & 0.561 & 0.439 & 0.551 & 0.237 & 0.816 & 0.961 & 0.653 & 0.722 & 0.730 & 0.277 \\
& \textbf{ProstT5}  & 0.799 & \textbf{0.744} & 0.597 & \underline{0.499} & \textit{0.615} & 0.291 & \textit{0.870} & 0.974 & 0.713 & 0.731 & 0.771 & 0.404 \\
& \textbf{MIF}      & 0.796 & 0.732 & 0.606 & 0.463 & 0.529 & 0.253 & 0.864 & \underline{0.984} & 0.719 & 0.714 & 0.675 & 0.337 \\
& \textbf{ProteinMPNN} & \underline{0.821} & \underline{0.742} & 0.601 & \textit{0.481} & 0.570 & \underline{0.310} & 0.865 & 0.980 & 0.719 & \textbf{0.751} & 0.765 & \textit{0.419} \\
& \textbf{Foldseek 3Di} & 0.800 & 0.728 & \textbf{0.630} & \textbf{0.508} & \textbf{0.635} & \textit{0.307} & \underline{0.873} & 0.959 & \underline{0.750} & \textbf{0.751} & \textit{0.797} & 0.416 \\
& \textbf{Amino Aseed}  & \textbf{0.830} & \underline{0.742} & \textit{0.619} & 0.453 & 0.599 & \textbf{0.318} & 0.869 & \textbf{0.993} & \textit{0.738} & 0.717 & \underline{0.822} & \underline{0.427} \\
\rowcolor{gray!15} & \textbf{\name} & \textit{0.809} & \underline{0.742} & \underline{0.623} & 0.473 & \underline{0.620} & 0.300 & \textbf{0.879} & \textbf{0.993} & \textbf{0.764} & \underline{0.736} & \textbf{0.854} & \textbf{0.473} \\
\bottomrule
\end{tabular}
\vspace{-3mm}
\end{table}

\subsection{Loops with no known Canonical Cluster}
\begin{table}[H]
\vspace{-5mm}  
\centering
\captionof{table}{Proportion of loops in SAbDab with no known \cite{kelow2022penultimate} canonical cluster with a cutoff of $\mathcal{D} = 0.47$ to cluster centroids.}
\vspace{-3mm}
\small
\begin{tabular}{lcc}
\toprule
\textbf{CDR} & \textbf{Heavy} & \textbf{Light} \\
\midrule
CDR1 & 0.130 & 0.112 \\
CDR2 & 0.098 & 0.187 \\
CDR3 & 0.763 & 0.192 \\
CDR4 & 0.037 & 0.062 \\
\bottomrule
\end{tabular}
\label{tab:proportion_noise}
\end{table}

\subsection{Visualization of \name Latent Space}
\begin{figure}[H]
  \centering
  \includegraphics[width=\textwidth]{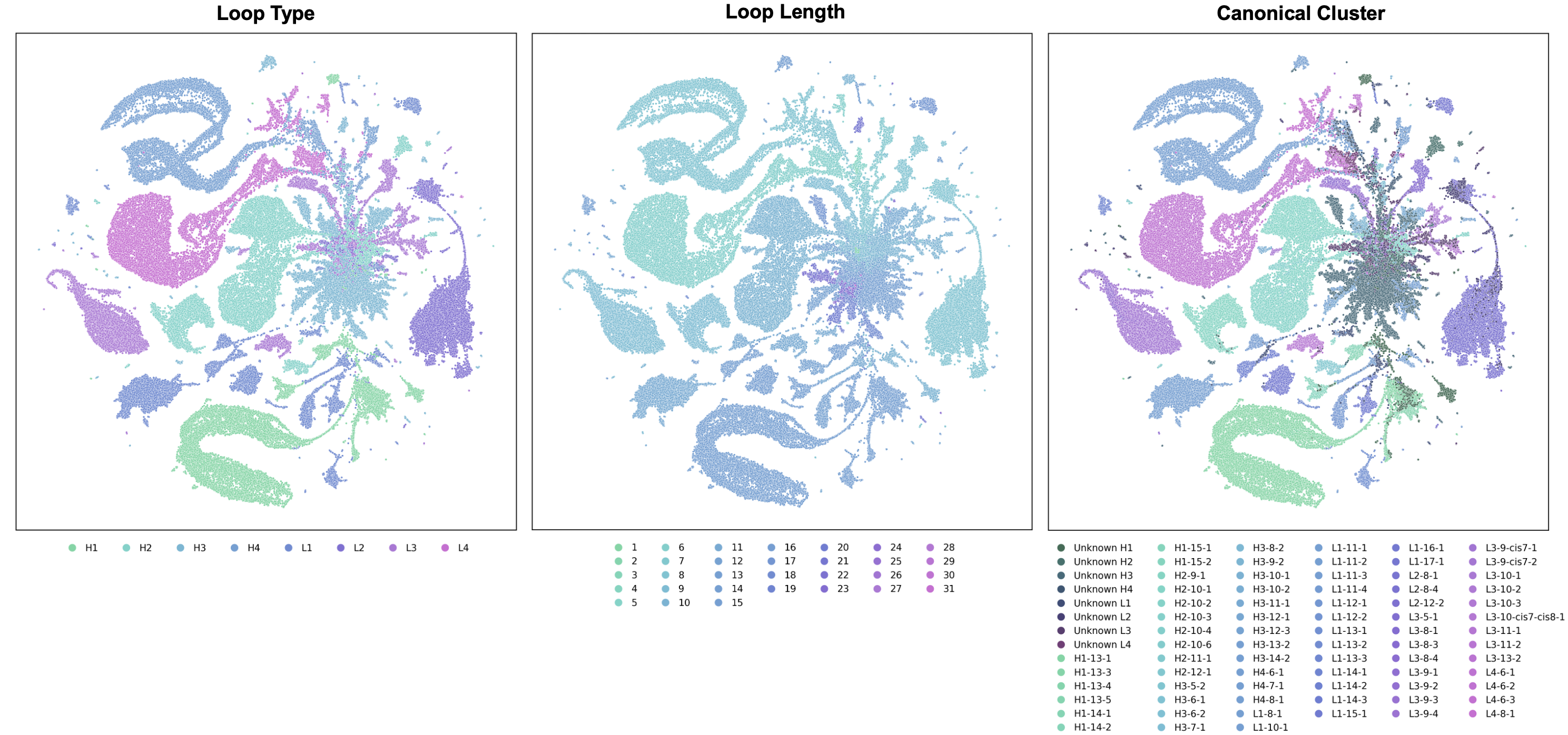}
  \vspace{-15pt}  
  \caption{UMAP of the \name latet space for loops in SAbDab. \textbf{Left} Loops colored by loop type. \textbf{Middle} Loops colored by loop length. \textbf{Right} Loops colored by their canonical cluster.}
  \label{fig:umap}
  \vspace{-10pt}  
\end{figure}

In Fig.~\ref{fig:umap} we visualize the \name token $\mathbf{t}$ for all loops in SAbDab across train, test, and validation datasets in 2D with Uniform Manifold Approximation and Projection (UMAP) \citep{mcinnes2018umap}. We observe in the UMAP that the embeddings are localized by their loop type, loop length, and canonical cluster. Among the CDRs, the H3 embeddings span the broadest region of the UMAP manifold, reflecting their markedly higher structural diversity.

\subsection{Recovery of the canonical clusters}
In Table~\ref{tab:sabdab_canonical_cluster_recovery} we report how well \name can recover \cite{kelow2022penultimate} clusters based on the \name quantized token, $\mathbf{\hat{t}}$. We compare the results with (1) \name with the dihedral angles masked out and only sequence input, and (2) \name with the sequence masked out and only dihedral angles input. \name performs best when dihedral angles are provided, though the dihedral angles can also be obtained through protein structure prediction models. Inference with only sequence input is suitable when encoding large libraries of antibody sequences, where structure prediction would be too computationally intensive. We observe that the performance of the sequence-only \name is very close to the \name model for most loop types except for the H3 and L3 loops. 

\begin{table}[H]
\centering
\caption{Average \name cluster purity ($\uparrow$) of \cite{north2011new} defined clusters of antibody CDRs across SAbDab.}
\vspace{1mm}
\label{tab:sabdab_canonical_cluster_recovery}
\vspace{-3mm}
\small
\begin{tabular}{lccc}
\toprule
\textbf{Loop Type} & \textbf{\name} & \textbf{\name sequence only} & \textbf{\name dihedral angles only} \\
\midrule
H1 & 0.894 & 0.880 & 0.898  \\
H2 & 0.900 & 0.875 & 0.914 \\
H3 & 0.754 & 0.537 & 0.725 \\
H4 & 0.983 & 0.996 & 0.979 \\
L1 & 0.880 & 0.841 & 0.867 \\
L2 & 0.975 & 0.991 & 0.976 \\
L3 & 0.831 & 0.771 & 0.812 \\
L4 & 0.930 & 0.928 & 0.914 \\
\bottomrule
\vspace{-4mm}
\end{tabular}
\end{table}

\subsection{Predicting Binding Affinity with \name Tokens}

In Table~\ref{tab:abibench_supp}, we show results of \almname compared to \plmname in predicting binding affinity for AbBiBench. \almname additionally includes multimodal residue tokens from \name which encode the dihedral angle backbone of the loop residues. We observe a drop in performance when these tokens are added. This is consistent with the lower performance of protein language models which use Foldseek 3Di tokens (SaProt and ProstT5 in Table~\ref{tab:abibench}). In the binding‑affinity benchmark, models must distinguish subtle differences among a small set of heavy‑chain variants. State‑of‑the‑art structure‑prediction models often miss these nuanced conformational changes \citep{pak2023using, buel2022can}, and the resulting errors propagate to protein language models that rely on residue‑level structural tokens.

\begin{table}[H]
\centering
\caption{Spearman correlation coefficient ($\uparrow$) for binding affinity prediction on AbBiBench for \plmname and \almname.}
\label{tab:abibench_supp}
\begin{tabular}{lcc}
\toprule
\textbf{Target} & \textbf{\plmname} & \textbf{\almname} \\
\midrule
1mlc & 0.616 (0.009) & 0.513 (0.020) \\
1n8z & 0.675 (0.025) & 0.556 (0.023) \\
2fjg & 0.713 (0.014) & 0.635 (0.015) \\
3gbn\_h1 & 0.948 (0.004) & 0.929 (0.005) \\
3gbn\_h9 & 0.962 (0.003) & 0.959 (0.002) \\
4fqi\_h1 & 0.921 (0.001) & 0.886 (0.001) \\
4fqi\_h3 & 0.971 (0.001) & 0.967 (0.001) \\
aayl49 & 0.625 (0.010) & 0.552 (0.007) \\
aayl49\_ML & 0.531 (0.007) & 0.493 (0.007) \\
aayl51 & 0.579 (0.011) & 0.545 (0.014) \\
\bottomrule
\end{tabular}
\end{table}

\subsection{Controllable Sampling of Antibody Loops}
\begin{figure}[H]
  \centering
  \vspace{-10pt}  
  \includegraphics[width=\textwidth]{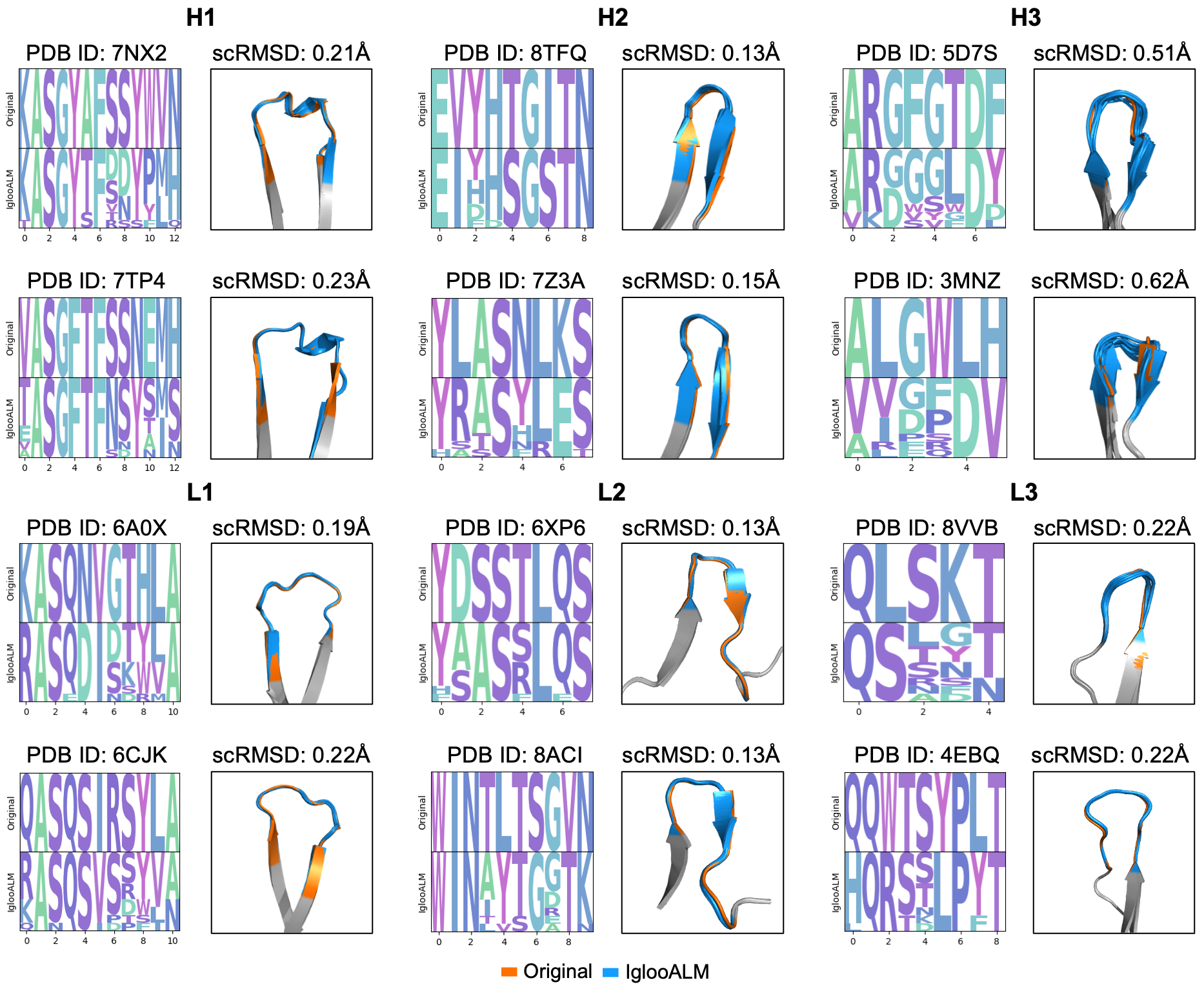}
  \vspace{-10pt}  
  \caption{
  Sequence logo of original and ten \almname sampled sequences of CDR loop regions at $\lambda=0.5$ and predicted structures. 
  }
  \label{fig:loop_rmsd_supp}
\end{figure}

In Figure~\ref{fig:loop_rmsd_supp} we present additional examples of \almname sampled sequences and their predicted structures. We present examples where the sampled loops have on average at most 60\% sequence identity with the original loop.

\end{document}